\documentclass[a4paper,fleqn]{cas-sc}

\usepackage[authoryear,longnamesfirst]{natbib}
\usepackage{bm}
\usepackage{booktabs}
\usepackage{multirow} 
\usepackage{blindtext}
\usepackage{makecell}
\usepackage{microtype}
\usepackage{array}
\usepackage{scalerel}
\usepackage{tabularx}
\newcolumntype{C}[1]{>{\centering\arraybackslash}p{#1}}
\usepackage{xspace}
\usepackage{adjustbox}

\def\tsc#1{\csdef{#1}{\textsc{\lowercase{#1}}\xspace}}
\tsc{WGM}
\tsc{QE}
\tsc{EP}
\tsc{PMS}
\tsc{BEC}
\tsc{DE}
\def\Tone{{T$_1$-w}\xspace}
\def\Ttwo{{T$_2$-w}\xspace}
\def\pq{$p_{\scaleto{q}{5pt}}$}
\def\pplus{\ensuremath{p_{\scaleto{\bm +}{4pt}}}}
\def\pminus{\ensuremath{p_{\scaleto{\bm -}{2pt}}^{\scaleto{(n)}{6pt}}}}
\def\xq{\ensuremath{x_{\scaleto{q}{5pt}}}}
\def\xplus{\ensuremath{x_{\scaleto{\bm{+}}{4pt}}}}
\def\xminus{\ensuremath{x_{\scaleto{\bm -}{2pt}}^{\scaleto{(m)}{6pt}}}}

\usepackage{fontawesome}
\usepackage[outline, auto]{contour}
\contourlength{0.50pt}
\newcommand{\tick}{\contour{Gray}{\textcolor{Green}{\faCheck}}}
\newcommand{\xmark}{\contour{Gray}{\textcolor{Maroon}{\faTimes}}}

\begin{document}
\let\WriteBookmarks\relax
\def\floatpagepagefraction{1}
\def\textpagefraction{.001}

\shorttitle{HACA3: A unified approach for multi-site MR image harmonization}

\shortauthors{Zuo~et~al.}

\title [mode = title]{HACA3: A unified approach for multi-site MR image harmonization}                      

%
\author[1,2]{Lianrui~Zuo}[orcid=0000-0002-5923-9097]
\cormark[1]
\ead{lr_zuo@jhu.edu}
\credit{Conceptualization of this study, Methodology, Data curation, Software, Writing --- Original draft, Experiments}
\address{{$^a$Department of Electrical and Computer Engineering, Johns Hopkins University}, {Baltimore}, {MD~21218}, {USA}}
\address{{$^b$Laboratory of Behavioral Neuroscience, National Institute on Aging, National Institutes of Health}, {Baltimore}, {MD~20892}, {USA}}
\address{{$^c$Department of Neurology, Johns Hopkins School of Medicine}, {Baltimore}, {MD~21287}, {USA}}
\address{{$^d$Department of Computer Science, Johns Hopkins University}, {Baltimore}, {MD~21218}, {USA}}
%
\author[1]{Yihao~Liu}
\credit{Conceptualization of this study, Methodology, Writing}
\author[1]{Yuan~Xue}
\credit{Conceptualization of this study, Methodology, Writing---review and editing}
\author[3]{Blake~E.~Dewey}
\credit{Conceptualization of this study, Methodology, Writing---review and editing}
\author[4]{Samuel~W.~Remedios}
\credit{Methodology, Writing---review and editing}
\author[1]{Savannah~P.~Hays}
\credit{Methodology, Writing---review and editing}
\author[2]{Murat~Bilgel}
\credit{Methodology, Writing---review and editing}
\author[3]{Ellen~M.~Mowry}
\credit{Methodology, Writing---review and editing, Supervision, Funding acquisition}
\author[3]{Scott~D.~Newsome}
\credit{Methodology, Writing---review and editing, Supervision, Funding acquisition}
\author[3]{Peter~A.~Calabresi}
\credit{Methodology, Writing---review and editing, Supervision, Funding acquisition }
\author[2]{Susan~M.~Resnick}
\credit{Resources, Methodology, Writing---review and editing, Supervision, Funding acquisition}
\author[1]{Jerry~L.~Prince}
\credit{Resources, Writing---review and editing, Supervision, Project administration, Funding acquisition.}
\author[1]{Aaron~Carass}
\credit{Conceptualization of this study, Writing---review and editing, Supervision}

\begin{abstract}
The lack of standardization is a prominent issue in magnetic resonance~(MR) imaging. 
This often causes undesired contrast variations in the acquired images due to differences in hardware and acquisition parameters. 
In recent years, image synthesis-based MR harmonization with disentanglement has been proposed to compensate for the undesired contrast variations. 
Despite the success of existing methods, we argue that three major improvements can be made. 
First, most existing methods are built upon the assumption that multi-contrast MR images of the same subject share the same anatomy. 
This assumption is questionable, since different MR contrasts are specialized to highlight different anatomical features.
Second, these methods often require a fixed set of MR contrasts for training~(e.g., both T1-weighted and T2-weighted images), limiting their applicability. 
Lastly, existing methods are generally sensitive to imaging artifacts. 
In this paper, we present Harmonization with Attention-based Contrast, Anatomy, and Artifact Awareness (HACA3), a novel approach to address these three issues. 
HACA3 incorporates an anatomy fusion module that accounts for the inherent anatomical differences between MR contrasts.
Furthermore, HACA3 is also robust to imaging artifacts and can be trained and applied to any set of MR contrasts. 
HACA3 is developed and evaluated on diverse MR datasets acquired from 21 sites with varying field strengths, scanner platforms, and acquisition protocols.
Experiments show that HACA3 achieves state-of-the-art performance under multiple image quality metrics. 
We also demonstrate the applicability and versatility of HACA3 on downstream tasks including white matter lesion segmentation and longitudinal volumetric analyses.
Code will be publicly available upon paper acceptance.
\end{abstract}




\maketitle

\section{Introduction}
\label{sec:introduction}
Magnetic resonance~(MR) imaging is a widely used and flexible imaging modality for studying the human brain. 
By modifying underlying pulse sequences, multiple MR tissue contrasts can be acquired in a single imaging session, revealing different tissue properties and  pathology~\citep{prince2006medical}.
For example, T$_1$-weighted~(\Tone) images typically show balanced soft tissue contrast between gray matter~(GM) and white matter~(WM). 
T$_2$-weighted~(\Ttwo) fluid-attenuated inversion recovery~(FLAIR) images can detect WM lesions~\citep{brown2014magnetic}. 
However, the flexibility of MR imaging also introduces drawbacks, most notably the lack of standardization between imaging studies. 
Changes in pulse sequences, imaging parameters, and scanner manufacturers often cause undesired contrast variations in acquired images. 
These contrast variations are frequently observed in multi-site and longitudinal studies, where acquiring images with identical protocols and platforms is challenging.
It has been shown that directly processing these images without compensating for  contrast variations can lead to biased and inconsistent measurements, also known as the domain shift problem~\citep{biberacher2016intra, he2020self, zuo2021unsupervised}. 

\begin{figure}[!tb]
    \centering
        \includegraphics[width=0.99\textwidth]{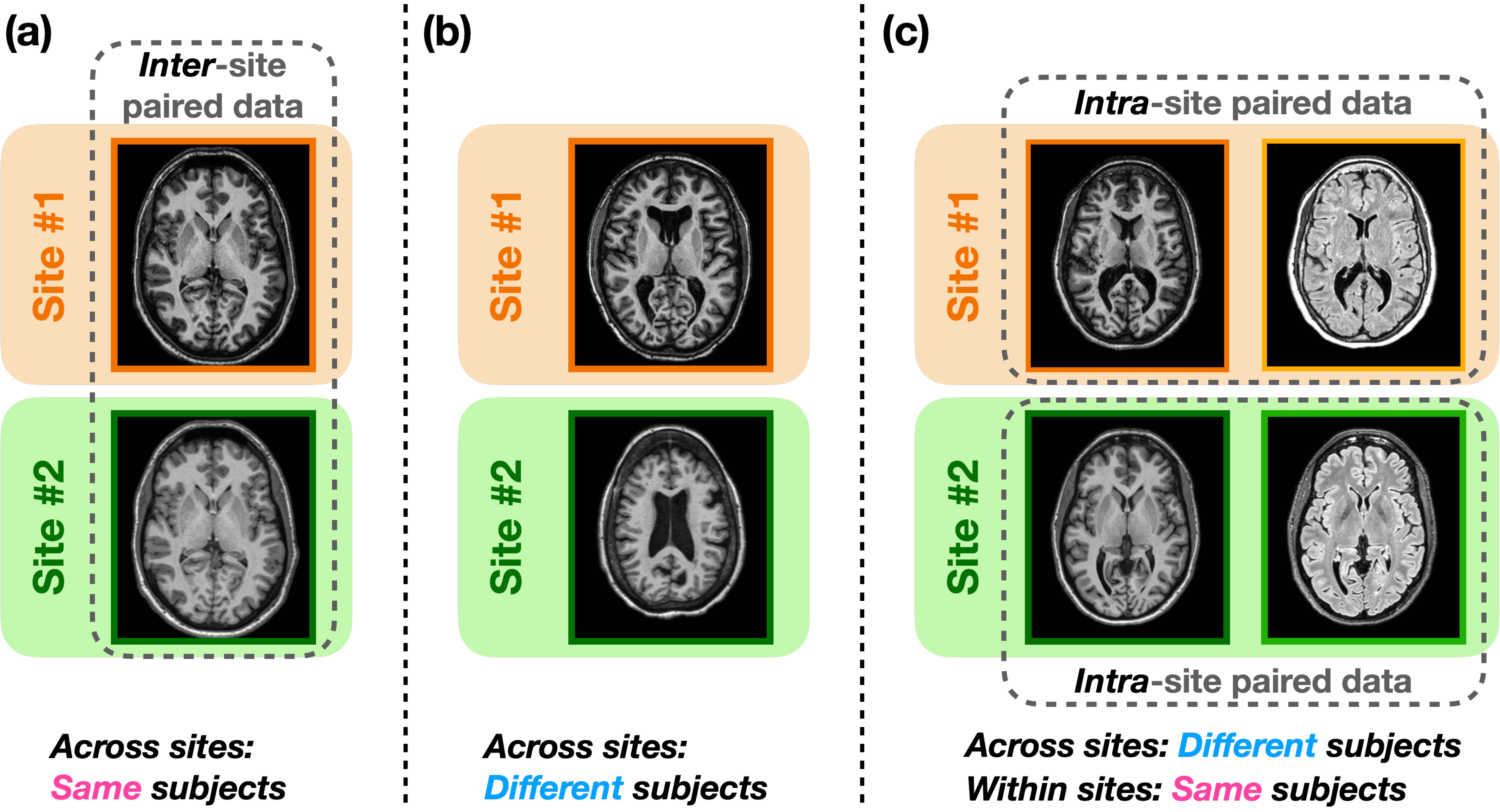}
    \caption{Training data required of the three types of harmonization methods. \textbf{(a)}~Supervised harmonization methods~\citep{dewey2019deepharmony,tian2022deep} require a sample group of subjects to be imaged across sites (i.e., inter-site paired data) for training. \textbf{(b)}~Unsupervised methods developed for natural image I2I~\citep{huang2018multimodal, liu2018unified, park2020contrastive, zhu2017unpaired} can be trained with different subjects across sites. \textbf{(c)}~Unsupervised harmonization methods with disentanglement~\citep{ouyang2021representation, zuo2021information, zuo2021unsupervised} utilize the routinely acquired intra-site paired data for training.}
    \label{fig:harmonization_methods}
\end{figure}

\textit{Harmonization through image synthesis alleviates domain shift.} In recent years, image synthesis-based MR harmonization techniques~\citep{beizaee2023harmonizing, dewey2019deepharmony, DEWEY2022217, gebre2023cross, liu2021style, zuo2021information, zuo2021unsupervised, zuo2022disentangle} have emerged to mitigate the lack of standardization in MR imaging.
These methods are a special type of image-to-image translation~(I2I)~\citep{huang2018multimodal, park2020contrastive, liu2023one, roy2013tmi, zhu2017unpaired, zuo2020synthesizing}, where the source and target images $x$ and $y$ come from different sites, such as two different \Tone images from separate sites.
In this context, we assume that images acquired with the same hardware and software come from the same (imaging) site.
These harmonization methods learn a function, $f(\cdot)$, that translates $x$ from a source site to a target site, i.e., $\hat{y} = f(x)$ while \textit{preserving the underlying anatomy}. 
Depending on the required training data, existing harmonization methods can be categorized into supervised and unsupervised methods.
Supervised harmonization methods~\citep{dewey2019deepharmony, tian2022deep} require a sample population to be imaged at multiple sites.
The acquired images across sites~(i.e., \textit{inter}-site paired data) are then used to train $f(\cdot)$, as shown in Fig.~\ref{fig:harmonization_methods}(a).
Although supervised harmonization generally exhibits superior performance due to the explicit voxel-level supervision provided by the inter-site paired data, their utility is limited to sites visited by traveling subjects.
Conversely, unsupervised harmonization methods do not require inter-site paired data, thereby offering broader applicability. 
Most existing unsupervised methods for natural image I2I, such as CycleGAN~\citep{zhu2017unpaired}, UNIT~\citep{liu2017unsupervised}, MUNIT~\citep{huang2018multimodal}, and CUT~\citep{park2020contrastive} can be used to achieve unsupervised harmonization by translating MR images across imaging sites, as shown in Fig.~\ref{fig:harmonization_methods}(b).
Even though cycle-consistency loss in anatomy is typically used in these methods to encourage the preservation of anatomical features during I2I, geometry shift remains a significant issue due to the absence of direct supervision on anatomy across sites. 
Recent studies have shown that the cycle-consistency constraint is insufficient for unsupervised I2I in medical imaging~\citep{gebre2023cross, yang2018unpaired, zuo2021unsupervised}.

\textit{A unique aspect of MR imaging motivates better unsupervised harmonization.} A 
distinctive feature of MR imaging is the routine acquisition of multi-contrast images of the same subject (i.e., \textit{intra}-site paired data) within a single imaging session to highlight different anatomical properties. 
For example, the publicly available IXI~\citep{IXI} dataset includes \Tone, \Ttwo, and proton density-weighted~(PD-w) images from different imaging sites. 
The OASIS3~\citep{LaMontagne} dataset has intra-site paired \Tone and \Ttwo images.
In recent years, unsupervised harmonization methods with disentanglement have been proposed to utilize intra-site paired data for improved harmonization. 
Figure~\ref{fig:harmonization_methods}(c) illustrates the training data used by these methods, where multi-contrast images of the same subject within each imaging site are employed.
The core concept is to disentangle anatomical and contrast~(i.e., acquisition related) information using \textit{intra}-site paired images during training, so that the anatomy information and a desired contrast can be recombined at test time to achieve \textit{inter}-site harmonization.
For instance, \citet{zuo2021information,zuo2021unsupervised} disentangled anatomical and contrast information given intra-site paired \Tone and \Ttwo images. 
In their work, disentanglement was achieved with adversarial training and a similarity loss, assuming that the intra-site paired images share the same anatomical information.
\citet{ouyang2021representation} learned disentangled anatomy and contrast representations based on intra-site paired data with a margin hinge loss.
The authors reported superior performance over existing unsupervised I2I methods such as CycleGAN~\citep{zhu2017unpaired}, due to supervision in geometry provided by the intra-site paired data. 

\begin{figure}[!tb]
    \centering
    \includegraphics[width=0.55\textwidth]{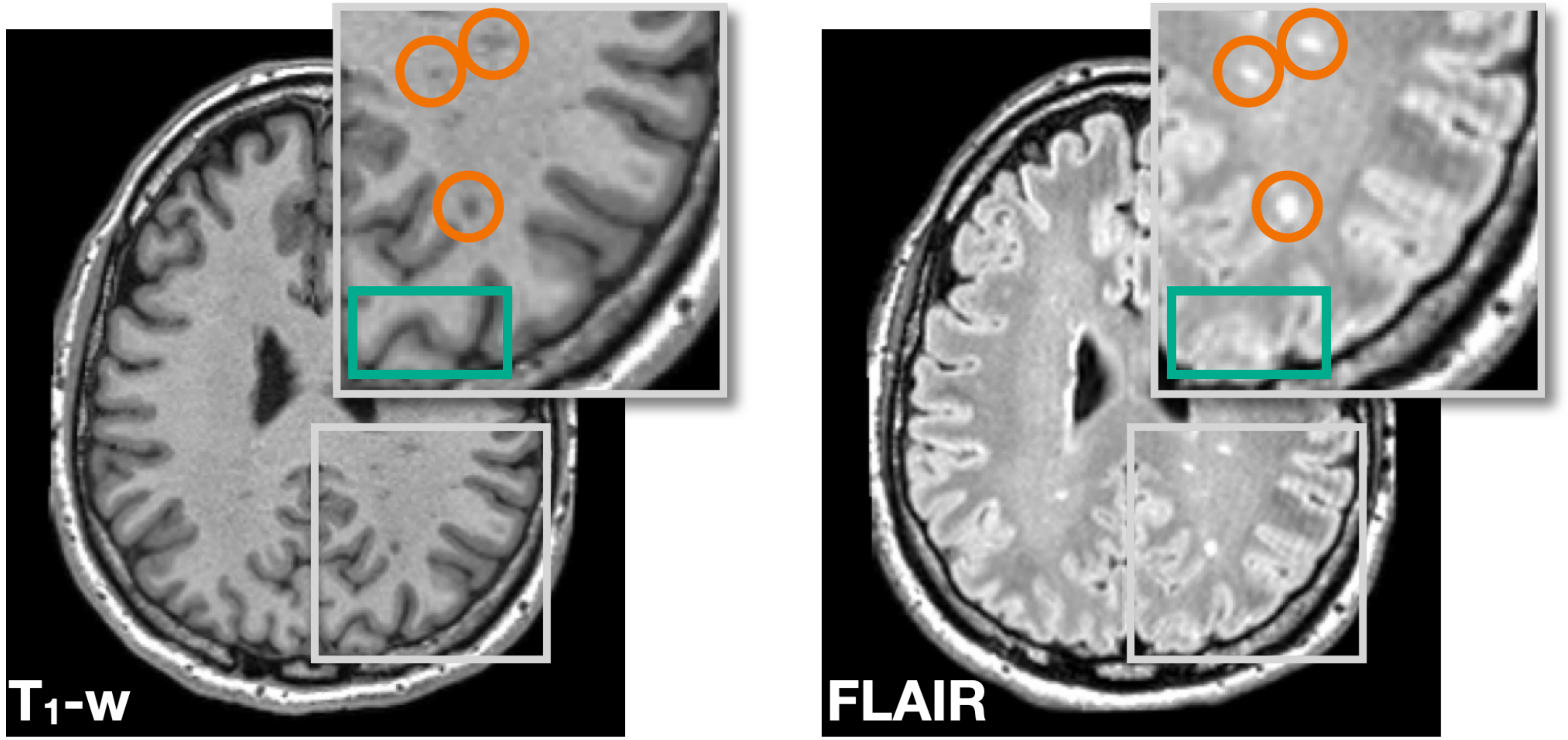}
    \caption{\Tone and FLAIR images of a multiple sclerosis~(MS) subject reveal slightly different anatomical features. The \Tone image shows better contrast between GM, WM, and cerebrospinal fluid (highlighted by the green box), while the FLAIR image shows clearer boundaries for the WM lesions (highlighted by the orange circles).}
    \label{fig:t1w_flair}
\end{figure}

\textit{However, current unsupervised harmonization methods miss an important consideration.} 
Most disentangling methods assume that intra-site paired images share \textit{identical} underlying anatomy while only differing in image contrast~\citep{chartsias2019disentangled, dewey2020disentangled, liu2022disentangled, zuo2021information, zuo2021unsupervised}.
This assumption is commonly used as an inductive bias, which is fundamental to learn disentanglement, according to~\citet{locatello2019challenging}.
However, an overlooked aspect is that different MR contrasts are specifically designed to better reveal different pathology, which implies that \underline{the commonly used assumption of identical anatomy is not strictly accurate in MR imaging.}
As shown in Fig.~\ref{fig:t1w_flair}, although the two images come from the same subject, different MR contrasts reveal slightly different anatomical information.
Specifically, the \Tone image shows better contrast between GM, WM, and cerebrospinal fluid (highlighted by the green box), while the FLAIR image shows clearer boundaries for the WM lesions (highlighted by the orange circles).
In this sense, these intra-site paired data are not perfect due to inherent anatomical differences between contrasts in MR imaging.
Recent work by~\citet{trauble2021disentangled} have both theoretically and practically identified trade-offs between disentanglement and the quality of synthetic images when using imperfect paired data during training.
Follow-up works in the medical domain by~\citet{ouyang2021representation} and~\citet{zuo2022disentangle} have reported on the negative impact of enforcing identical anatomies of intra-site paired data during image synthesis.

Several unresolved problems persist with training a harmonization model that respects the anatomical differences between MR images with different acquisitions.
First, the observable anatomy of intra-site paired images should be considered different, necessitating a new inductive bias to achieve disentanglement. 
Second, the MR tissue contrasts from the source site have an impact on harmonization. 
Ideally, the model should understand MR tissue contrast and choose an appropriate combination of contrasts to produce better harmonization.
Third, imaging artifacts and missing contrasts should be handled to improve robustness and applicability.

In this paper, we propose a novel \underline{h}armonization approach to address these three issues with \underline{a}ttention-based \underline{c}ontrast, \underline{a}natomy, and \underline{a}rtifact \underline{a}wareness (HACA3). 
The contributions of the paper are as follows:
\begin{itemize}
    \item We challenge the common assumption of identical anatomy for MR disentanglement and propose a new inductive bias to learn disentanglement from MR images. As a result, HACA3 respects the inherent anatomy difference between MR contrasts. 
    \item We design a novel contrast and artifact attention mechanism to produce an optimal harmonized image based on the contrast and artifact information of each input image.
    \item HACA3 can be trained and applied to any set of MR contrasts by using a special design to handle missing contrasts.
\end{itemize}
HACA3 outperforms existing harmonization and I2I methods according to multiple image quality metrics. 
We use diverse MR datasets to demonstrate the broad applicability of HACA3 in downstream tasks including WM lesion segmentation and longitudinal volumetric analyses.

\section{Methods}
\label{sec:methods}
\subsection{General framework}
\begin{figure}[!t]
    \centering
    \includegraphics[width=0.99\textwidth]{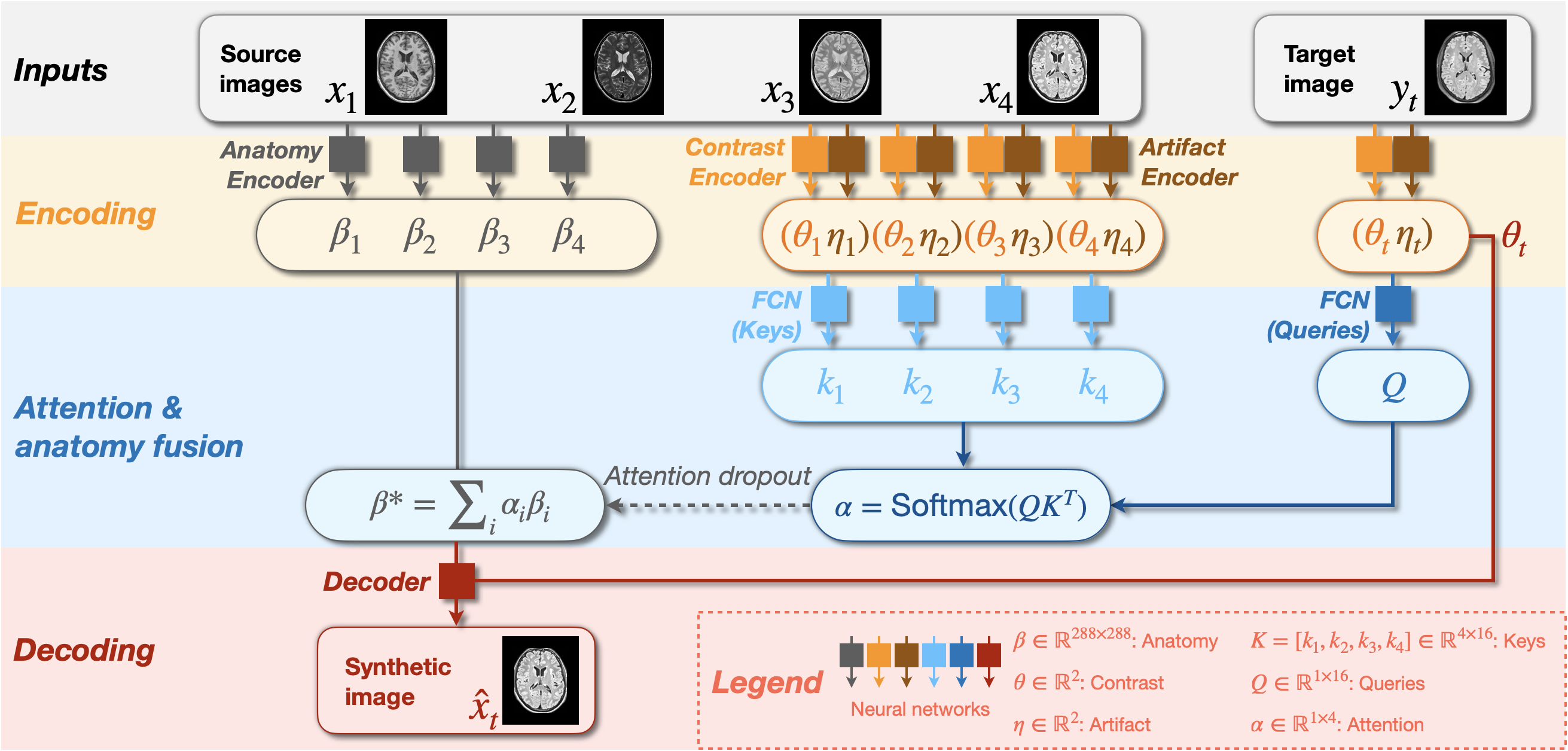}
    \caption{Schematic framework of HACA3. Networks with the same color share weights. Synthetic image $\hat{x}_t$ has the same contrast as the target image $y_t$ while preserving the anatomy from source images. Networks to process keys and queries are both fully connected networks (FCNs).}
    \label{fig:framework}
\end{figure}

HACA3 follows an ``encoder--attention--decoder'' structure. 
In constrast to existing frameworks that disentangle anatomy and contrast~\citep{dewey2020disentangled, liu2022disentangled, ouyang2021representation, zuo2021unsupervised}, HACA3 has an additional encoder---the artifact encoder---to assess the extent of artifacts present in the input MR images.
Additionally, we introduce an attention module that analyzes the learned representations of contrast, anatomy, and artifacts to inform the decoder for better harmonization.
Figure~\ref{fig:framework} shows the schematic framework of HACA3, which comprises three major components: 1)~encoding, 2)~anatomy fusion with attention, and 3)~decoding.
In this section, we provide an overview of HACA3's general ideas. 
We offer detailed explanations of the encoding and attention components in Secs.~\ref{sec:disentangle} and~\ref{sec:attention}, respectively.
Implementation details, including network architectures and training losses, are described in Secs.~\ref{sec:network} and~\ref{sec:losses}.

During training, HACA3 encodes the intra-site \Tone, \Ttwo, PD-w, and FLAIR images~($x_1$, $x_2$, $x_3$, and $x_4$, respectively) of the same subject into anatomy representations $\beta$, contrast representations $\theta$, and artifact representations $\eta$ using three corresponding encoders $E_\beta(\cdot)$, $E_\theta(\cdot)$, and $E_\eta(\cdot)$, respectively. 
It is necessary to note that \underline{HACA3 does not require all four contrasts for training}; it can be trained with any combination of MR tissue contrasts, as we describe in Sec.~\ref{sec:attention}.
Contrast and artifact representations ($\theta_t$ and $\eta_t$) of the target image $y_t$ are also calculated during encoding.
Following~\citet{chartsias2019disentangled, dewey2020disentangled, ouyang2021representation, zuo2021information, zuo2022disentangle}, HACA3 conducts intra-site I2I~(e.g., intra-site \Tone to \Ttwo synthesis) with disentangled representations $\theta$ and $\beta$ during training.
At test time, $\theta$ and $\beta$ from different sites are recombined to achieve inter-site harmonization.
The anatomy representation $\beta$ has the same spatial dimension as images $x$ with five distinct intensity levels, calculated from a five-channel one-hot encoded map using Gumbel softmax.
This choice of anatomy representation has been explored and validated in multiple disentangling works~\citep{chartsias2019disentangled, dewey2020disentangled, liu2020variational, zuo2021unsupervised, zuo2022disentangle}.
The contrast representation $\theta$ and artifact representation $\eta$ are two-dimensional variables~(i.e., $\theta, \eta \in \mathbb{R}^2$).
HACA3 then employs an attention module (we describe in Sec.~\ref{sec:attention}) to process the learned representations $\theta$ and $\eta$ and find the optimal anatomy representation $\beta^*$ for the target image $y_t$. 
The decoder subsequently recombines $\beta^*$ and $\theta_t$ to generate a harmonized image $\hat{x}_t$ with the desired contrast as $y_t$ while preserving the anatomy from the source images.

\subsection{Encoding: contrast, anatomy, and artifacts}
\label{sec:disentangle}
\subsubsection{A new inductive bias to disentangle anatomy and contrast}
\begin{figure}[!tb]
    \centering
    \includegraphics[width=0.9\textwidth]{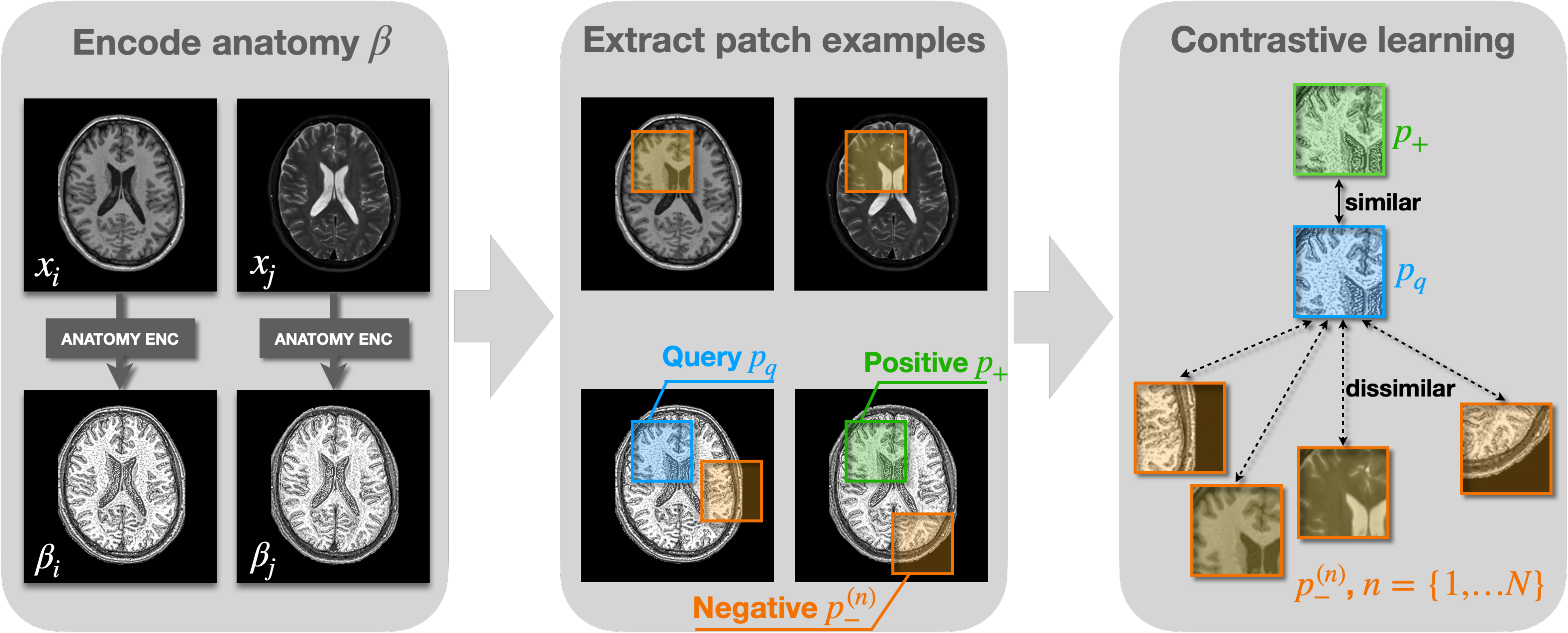}
    \caption{Learning anatomical representations $\beta$ with contrastive learning. \pq, \pplus, and \pminus{} are query patch, positive patch, and negative patches, respectively. In previous works, \pq{} is encouraged to be equal to \pplus{}. In our work, \pq{} is encouraged to be more similar to \pplus{} than to \pminus{} using Eq.~\ref{eq:contrastive_loss}, where $n=\{1,\dots,N\}$ and $N$ is the number of negative patches.}
    \label{fig:contrastive_learning}
\end{figure}

We introduce a novel way to disentangle anatomy and contrast while respecting the natural anatomy differences between MR contrasts.
The core concept of our anatomy encoder is based on contrastive learning~\citep{park2020contrastive}, which learns discriminative features from query, positive, and negative examples.
Here, the query, positive, and negative examples are small image patches denoted as \pq{}, \pplus{}, and $p_{\scaleto{\bm -}{2pt}}$, respectively. 
As shown in Fig.~\ref{fig:contrastive_learning}, intra-site paired images of different MR contrasts are individually processed by the anatomy encoder to learn anatomical representations, where $i, j \in \{1,2,3,4\}$ ($i\neq j$) are randomly selected contrasts.
The query patch, \pq, is selected at a random location of $\beta_i$, i.e., the anatomical representation of contrast $i$.
The positive patches \pplus{} are selected at the corresponding locations of $\beta_j$, where $j \neq i$.
Negative patches \pminus{} are sampled at the same locations as \pq{} from the original images as well as random locations from the learned $\beta$'s.
Previous works~\citep{chartsias2019disentangled, dewey2020disentangled, liu2022disentangled, zuo2021information, zuo2021unsupervised} have attempted to enforce identical anatomical representations between different MR contrasts. 
In other words, the query patch \pq{} equals to the positive patch \pplus{} at every location.
However, as we discussed in Sec.~\ref{sec:introduction}, this assumption is not entirely true. 
In our work, instead of enforcing \pq{} to be identical to \pplus{}, we encourage \pq{} to be more similar to \pplus{} than to the \pminus's using
\begin{equation}
\label{eq:contrastive_loss}
    \mathcal{L}_{\mbox{\tiny{C}}}(p_{\scaleto{q}{4pt}}, p_{\scaleto{\bm +}{4pt}}, p_{\scaleto{\bm  -}{2pt}}^{\scaleto{(n)}{6pt}}) = 
    -\log \left[ \frac{\exp( p_{\scaleto{q}{4pt}} \cdot p_{\scaleto{\bm +}{4pt}})}{ \exp(p_{\scaleto{q}{4pt}} \cdot p_{\scaleto{\bm +}{4pt}}) + \frac{1}{N}\sum_{n=1}^N \exp(p_{\scaleto{q}{4pt}} \cdot p_{\scaleto{\bm -}{2pt}}^{\scaleto{(n)}{6pt}})} \right].
\end{equation}
Our inductive bias for disentanglement is that no matter how similar \pq{} and \pminus's are, the positive patches \pplus{} should always be more similar to \pq, but not necessarily identical. 
The intuition is that \pq{} and \pplus{} are representations of the same subject, while \pminus's either represent different anatomical information or the same subject with weighted contrasts. 
Equation~\ref{eq:contrastive_loss} essentially encourages contrast information to be removed from $\beta$. 
Because our decoder takes both $\beta$ and $\theta$ as direct inputs to generate a harmonized image during training, contrast information is pushed to the $\theta$ branch, which we adopt from~\cite{zuo2021unsupervised}.

\subsubsection{Learning representations of artifacts}
\label{sec:artifacts}
\begin{figure}[!tb]
    \centering
    \includegraphics[width=0.9\textwidth]{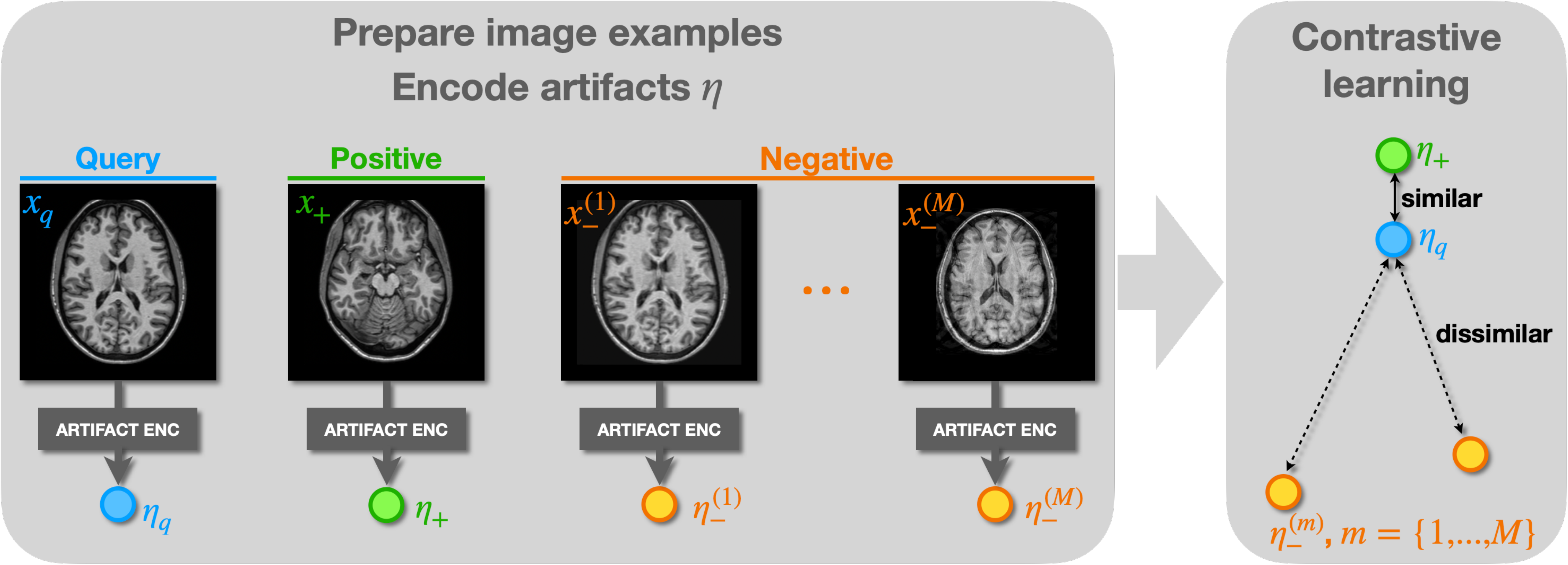}
    \caption{Learning artifact representations $\eta \in \mathbb{R}^2$ with contrastive learning. \xq{} and \xplus{} are assumed to have the same level of artifacts, while \xq{} and \xminus{} have different levels of artifacts. The contrastive loss is applied to encourage $\eta$ to preserve this relationship.}
    \label{fig:artifact_encoder}
\end{figure}

Our artifact encoder~$E_\eta(\cdot)$ is designed to capture imaging artifacts that commonly occur in MR images and can negatively affect harmonization performance.
By learning artifact representations $\eta$ from source MR images $x$, i.e., $\eta = E_\eta(x)$, the harmonization model is informed to avoid using images with high levels of artifacts during application.
$E_\eta(\cdot)$, based on~\citep{zuo2023latent}, is also trained with contrastive learning, with query, positive, and negative examples being MR image slices denoted as \xq, \xplus, and \xminus, respectively. 
We prepare \xq{} and \xplus{} by selecting 2D image slices from the same 3D MR volume, assuming they have \textit{similar} levels of artifact. 
Negative examples \xminus{} are prepared in two ways: 1)~by augmenting \xq{} with simulated artifacts, such as motion and noise, and 2)~by selecting 2D image slices from different volumes than \xq. 
In both cases, we assume \xminus's and \xq{} have \textit{different} levels of artifact.
Since both simulated and real MR images are used as negative examples, $E(\eta)$ after training captures various artifacts beyond just motion and noise, as we demonstrated in our previous work~\cite{zuo2023latent}.
As shown in Fig.~\ref{fig:artifact_encoder}, query, positive, and negative images are processed by our artifact encoder $E_\eta(\cdot)$ to calculate the corresponding artifact representations $\eta$.
The final loss to train $E_\eta(\cdot)$ is given by the contrastive loss $\mathcal{L}_{\mbox{\tiny{C}}}(\eta_{\scaleto{q}{4pt}}, \eta_{\scaleto{\bm +}{4pt}}, \eta_{\scaleto{\bm -}{2pt}}^{\scaleto{(m)}{6pt}})$ in Eq.~\ref{eq:contrastive_loss}, where $m = \{1,\dots, M\}$ and $M$ is the total number of negative example images.

\subsection{Decoding with attention}
\label{sec:attention}
Given that $\beta_i$'s~($i=\{1,2,3,4\}$) from different images of the same subject should be similar but not necessarily identical, the choice of $\beta_i$ during decoding is crucial for successful harmonization.
When harmonizing an MR contrast from a target site, it is intuitive to choose $\beta$ of the same contrast from the source site since similar pulse sequences usually reveal similar underlying anatomical information. 
However, this approach may not always be optimal when dealing with imaging artifacts and poor image quality.
Alternatively, one can calculate $\beta$'s from all the available contrasts of the source images, which provides increased robustness against imaging artifacts and poor image quality.
Previous works~\citep{chartsias2019disentangled, ouyang2021representation, zuo2021information} have used either of the two ways separately, but HACA3 takes a step further by combining the advantages of both methods.
Specifically, we propose a novel attention mechanism that takes both contrast and artifact into consideration when fusing anatomy from multiple source images.
To do so, we use fully connected networks (FCNs) to learn keys $K = [k_1, k_2, k_3, k_4]$ and queries $Q$~\citep{vaswani2017attention}, from the encoded $\theta$ and $\eta$ of both source and target images, as shown in Fig.~\ref{fig:framework}.
We then obtain attentions $\alpha \in \mathbb{R}^4$ by measuring the similarity between $K$ and $Q$~\citep{vaswani2017attention}.
The learned attentions highlight the corresponding source images that have similar contrast and image quality as the target image $y_t$ used for harmonization.
Here, we assume that the target image $y_t$ has good image quality.
The optimal anatomical representation, $\beta^*$, is then obtained by conducting a weighted average with attention, i.e., $\beta^* = \sum_{i=1}^4 \alpha_i \beta_i$.
Finally, the decoder combines both $\beta^*$ and $\theta_t$ to generate a synthetic image~$\hat{x}_t$.

To enable HACA3 to handle an arbitrary number of MR contrasts during training, we introduce an attention dropout mechanism. 
When there are missing contrasts during training, the corresponding $\alpha_i$ is set to zero and the remaining $\alpha_i$'s are renormalized.
This ensures that $\sum_{i=1}^4 \alpha_i = 1$ and $\beta$ of the missing contrasts will not be selected while calculating $\beta^*$. 
Even when all four contrasts are available during training, one or more of the $\alpha_i$'s still have a chance to be randomly dropped out (set to zero), and the remaining $\alpha_i$'s are renormalized accordingly.
During application, HACA3 handles missing contrasts in source images in a similar manner by setting the corresponding $\alpha_i$ to zero.

\subsection{Network architectures}
\label{sec:network}
\begin{figure}[!tb]
    \centering
    \includegraphics[width=0.99\textwidth]{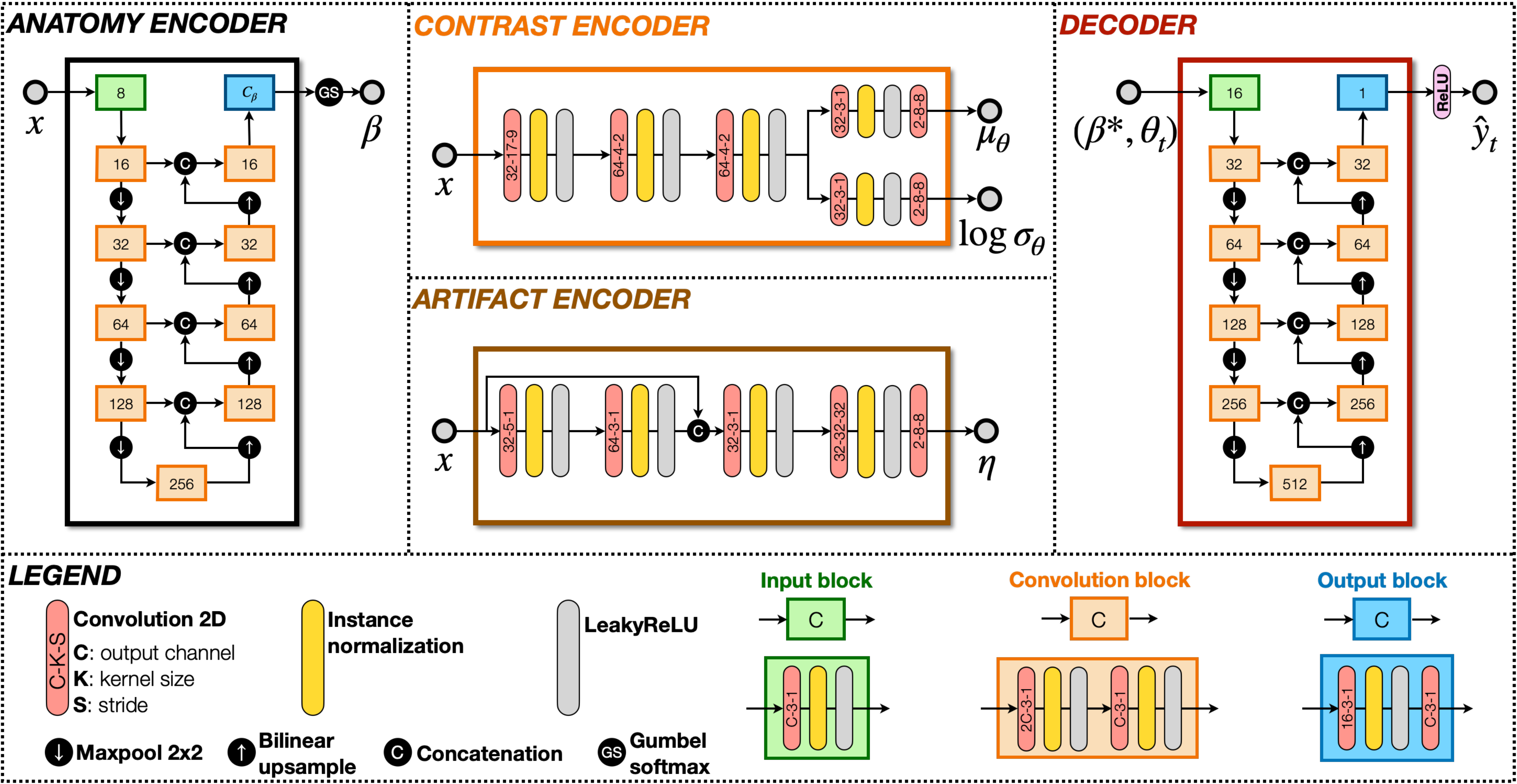}
    \caption{Network architectures of HACA3. The anatomy encoder and decoder are both U-Nets. }
    \label{fig:network_architecture}
\end{figure}
Network architectures are shown in Fig.~\ref{fig:network_architecture}.
Our anatomy encoder and decoder are both U-Nets~\citep{ronneberger2015miccai} with four downsampling layers. 
The decoder has double the channels of the anatomy encoder, because we believe it needs larger network capacity to generate various MR contrasts.
The contrast encoder is a fully convolutional network with four ``Convolution--InstanceNorm--LeakyReLU'' modules. 
The first convolutional kernel of our contrast encoder has a large kernel size. 
Because we believe contrast information of an MR image should be relatively global, using a large convolutional kernel to reduce the spatial dimension can help the model capture contrast information.
Our artifact encoder has a DenseNet structure with four convolutional layers adopted from~\cite{zuo2023latent}. 

\subsection{Implementation details and loss functions}
\label{sec:losses}
The framework of HACA3 is a conditional variational autoencoder~(CVAE) with $\theta$ being the latent variable.
The condition of the CVAE is $\beta^*$.
The CVAE loss to train HACA3 is given by 
\begin{equation}
    \mathcal{L}_{\mbox{\tiny{CVAE}}} = |\hat{x}_t - y_t|_1 + \lambda_1 \mathcal{D}_{\mbox{\tiny{KL}}}[p(\theta|y_t) || p(\theta)],
\end{equation}
where $\mathcal{D}_{\mbox{\tiny{KL}}}$ is the KL divergence.
To further regularize HACA3, synthetic image $\hat{x}$ is then reanalyzed by the encoders $E_\theta(\cdot)$ and $E_\eta(\cdot)$ and a cycle consistency loss is calculated, i.e., $\mathcal{L}_{\mbox{\tiny{cyc}}} = |E_\theta(\hat{x}_t)- \theta_t|_1 + |E_\eta(\hat{x}_t) - \eta_t|_1 $.
The overall loss to train HACA3 includes $\mathcal{L}_{\mbox{\tiny{CVAE}}}$, contrastive losses for anatomy and artifact encoders (see Eq.~\ref{eq:contrastive_loss}), and $\mathcal{L}_{\mbox{\tiny{cyc}}}$, i.e., 
\begin{equation}
    \mathcal{L}_{\mbox{\tiny{total}}} = |\hat{x}_t - y_t|_1 + \lambda_1 \mathcal{D}_{\mbox{\tiny{KL}}}[p(\theta|y_t) || p(\theta)] + \lambda_2 \mathcal{L}_{\mbox{\tiny{C}}}(p_{\scaleto{q}{4pt}}, p_{\scaleto{\bm +}{4pt}}, p_{\scaleto{\bm -}{2pt}}^{\scaleto{(n)}{6pt}}) + \lambda_3 \mathcal{L}_{\mbox{\tiny{C}}}(\eta_{\scaleto{q}{4pt}}, \eta_{\scaleto{\bm +}{4pt}}, \eta_{\scaleto{\bm -}{2pt}}^{\scaleto{(m)}{6pt}}) + \lambda_4 \mathcal{L}_{\mbox{\tiny{cyc}}},
\end{equation}
where $\lambda$'s are hyperparameters. 
In our implementation, $\lambda_1$ through $\lambda_4$ are $10^{-5}$, $0.1$, $0.1$, and $0.1$, respectively. 
Except for the KL divergence loss, the other loss terms have approximately equal magnitude after weighting by the $\lambda$'s. 
The KL divergence loss is lightly penalized in training, because previous works~\citep{higgins2017beta} have reported that when the KL divergence loss in a VAE model is heavily penalized, synthetic images tend to be blurry.
We also want to keep the KL divergence term, since this allows HACA3 to generate MR images with various contrasts by sampling $\theta$ space.
Dropping the KL divergence term will make HACA3 a conditional autoencoder, thus losing the ability to do variational sampling.

During training, target image $y_t$ is first randomly selected from intra-site paired images $x_1$ to $x_4$. 
In this case, HACA3 is trained to conduct {intra-site I2I} with disentanglement.
We also select $y_t$ from a different site than the source images during training. 
In this case, only $\mathcal{L}_{\mbox{\tiny{cyc}}}$ is calculated to train the attention module with {inter-site I2I}. 
Our code will be publicly available upon paper acceptance.

\section{Experiments and Results}
\subsection{Datasets and preprocessing}
\label{sec:data}
\begin{table}[!tb]
    \centering
    \caption{MR images acquired from $21$ sites were used to train and evaluate HACA3. Out of the $21$ sites, $11$ are publicly available \citep{IXI,LaMontagne,resnick2000one,carass2017longitudinal}. Magnetic field strengths are reported in teslas.}
    \resizebox{0.99\columnwidth}{!}{
    \begin{tabular}{C{0.14\columnwidth} C{0.06\columnwidth} C{0.06\columnwidth} C{0.06\columnwidth} C{0.06\columnwidth} C{0.06\columnwidth} C{0.06\columnwidth} C{0.06\columnwidth} C{0.06\columnwidth} C{0.06\columnwidth} C{0.06\columnwidth} C{0.06\columnwidth} }    
    \toprule 
        {\bf Site ID} & {$S_{1}$} & {$S_{2}$} & {$S_{3}$} & {$S_{4}$} & {$S_{5}$} & {$S_{6}$} & {$S_{7}$} & {$S_{8}$} & {$S_{9}$} & {$S_{10}$} &{$S_{11}$} \\
    \cmidrule(lr){1-12}
    {\bf Open data} & \tick & \tick & \tick & \tick & \tick & \tick & \tick 
                & \tick & \tick & \tick & \tick \\
    \cmidrule(lr){2-12}
    {\bf Manufacturer} & {Philips} & {Philips} & {Siemens} & {Siemens} & {Siemens} & {Siemens} & {Philips}  
                   & {Philips} & {Philips} & {Philips} & {Philips} \\ 
    \cmidrule(lr){2-12}
    {\bf {Field}} & {$1.5$} & {$3.0$} & {$3.0$} & {$3.0$} & {$3.0$} & {$1.5$} & {$1.5$} 
            & {$3.0$} & {$3.0$} & {$3.0$} & {$3.0$} \\
    \cmidrule(lr){2-12}
    \textbf{Population} & Healthy & Healthy & Healthy & Healthy & Healthy & Healthy & Healthy & Healthy & Healthy & Healthy & MS \\
    \cmidrule(lr){1-12}
    {\bf T$_{\boldsymbol{1}}$-w} 
            & \tick & \tick & \tick & \tick & \tick & \tick & \tick  
            & \tick & \tick & \tick & \tick \\
    \cmidrule(lr){2-12}
    {\bf T$_{\boldsymbol{2}}$-w} 
            & \tick & \tick & \tick & \tick & \tick & \tick & \tick  
            & \tick & \tick & \tick & \tick \\
    \cmidrule(lr){2-12}
    {\bf PD-w}  
            & \tick & \tick & \xmark & \xmark & \xmark & \xmark & \tick  
            & \tick & \tick & \tick & \tick \\
    \cmidrule(lr){2-12}
    {\bf FLAIR}  
            & \xmark & \xmark & \xmark & \xmark & \xmark & \xmark & \tick  
            & \tick & \tick & \tick & \tick \\
    \midrule \midrule
    {\bf Site ID} &{$S_{12}$} &{$S_{13}$} &{$S_{14}$} & {$S_{15}$} &{$S_{16}$} &{$S_{17}$} &{$S_{18}$} &{$S_{19}$} &{$S_{20}$} &{$S_{21}$} \\
    \cmidrule(lr){1-11}
    {\bf Open data} & \xmark & \xmark & \xmark & \xmark & \xmark & \xmark & \xmark & \xmark & \xmark & \xmark \\
    \cmidrule(lr){2-11}
    {\bf Manufacturer} & {Philips} & {Siemens} & {GE} & {Siemens} & {GE} & {Philips} & {Siemens} & {Siemens} & {Siemens} & {Siemens} \\
    \cmidrule(lr){2-11}
    {\bf Field} & {$3.0$} & {$3.0$} & {$1.5$} & {$3.0$} & {$3.0$} & {$3.0$} & {$3.0$} & {$3.0$} & {$1.5$} & {$3.0$} \\
    \cmidrule(lr){2-11}
    \textbf{Population} & MS & MS & MS & MS & MS & MS & MS & MS & MS & MS \\
    \cmidrule(lr){1-11}
    {\bf T$_{\boldsymbol{1}}$-w} & \tick & \tick & \tick & \tick & \tick & \tick & \tick & \tick & \tick & \tick \\
    \cmidrule(lr){2-11}
    {\bf T$_{\boldsymbol{2}}$-w} & \tick & \tick & \tick & \tick & \tick & \tick & \tick & \tick & \tick & \tick \\
    \cmidrule(lr){2-11}
    {\bf PD-w} & \tick & \tick & \xmark & \tick & \xmark & \tick & \xmark & \tick & \xmark & \tick \\
    \cmidrule(lr){2-11}
    {\bf FLAIR} & \tick & \tick & \tick & \tick & \xmark & \tick & \tick & \tick & \tick & \tick \\
    \bottomrule
    \end{tabular}}
    \label{tab:dataset}
\end{table}

As we show in Table~\ref{tab:dataset}, HACA3 is developed and evaluated with highly variable MR datasets acquired from $21$ sites, including healthy subjects~(Sites~$S_1$ to $S_{10}$) and subjects with multiple sclerosis~(MS)~(Sites~$S_{11}$ to $S_{21}$). 
Out of the 21 sites we used in our training and evaluation, sites $S_{13}$ to $S_{21}$ are clinical centers and have more variability in image acquisition parameters.
For sites $S_{13}$ to $S_{21}$, a small percentage of images acquired from the same site may have different acquisition parameters, leading to different image contrasts.

All images were preprocessed with inhomogeneity correction~\citep{tustison2010n4itk}, super-resolution for 2D acquired images~\citep{zhao2019mri, zhao2020smore}, registration to an MNI atlas with $0.8$mm$^3$ resolution, and a WM peak normalization~\citep{reinhold2019evaluating}.
For each site, ten training and two validation subjects were selected, each with two to four MR contrasts depending on availability. 
HACA3 was trained with 2D axial, coronal, and sagittal slices extracted from each 3D MR volume. 
We adopt the model introduced in~\citet{zuo2021unsupervised} to combine multi-orientation 2D slices into a 3D volume as our final harmonization result. 
Specifically, we use a 3D convolutional network that takes stacked 2D slices from axial, coronal, and sagittal orientations as input and generates a final 3D volume as output.

\subsection{Exploring the latent contrast, anatomy, and artifact space}
\label{sec:exploring_latent_space}
\begin{figure}[!tb]
    \centering
    \includegraphics[width=0.99\textwidth]{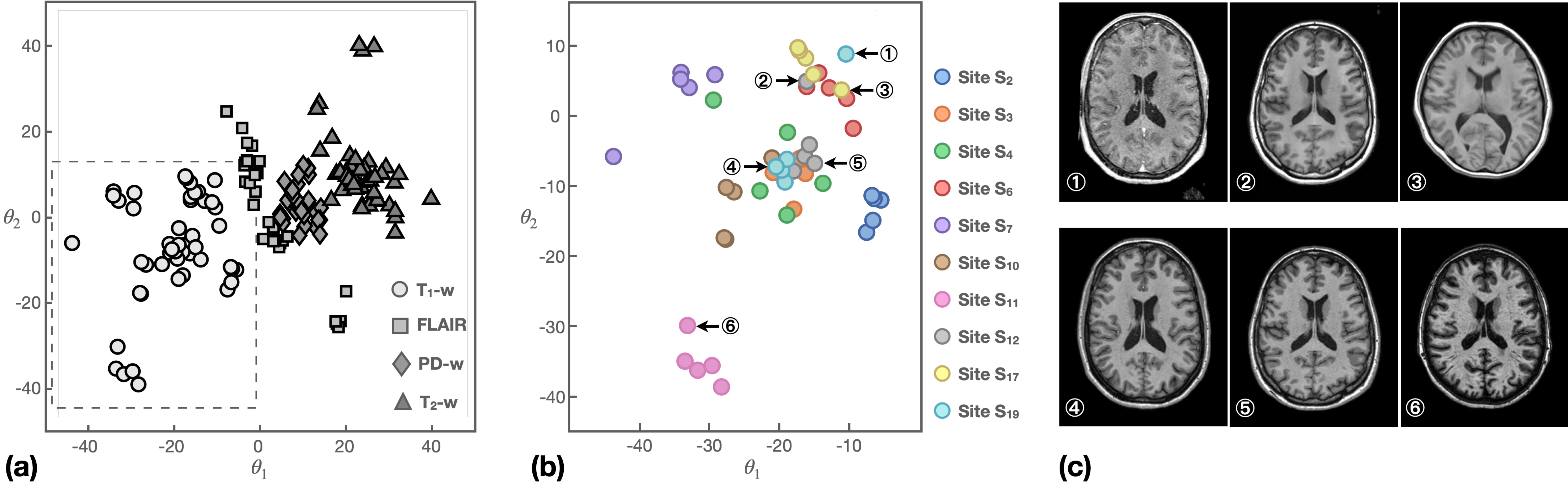}
    \caption{Contrast representations $\theta$ of 10 representative sites. \textbf{(a)}  $\theta$'s of \Tone, \Ttwo, PD-w, and FLAIR images. \textbf{(b)} $\theta$'s of \Tone images from different sites. Circled numbers show $\theta$ values of six representative images. Corresponding MR images are shown in \textbf{(c)}.}
    \label{fig:theta_space}
\end{figure}

The contrast encoder $E_\theta(\cdot)$ in HACA3 captures acquisition-related information from MR images. 
After training, we expect the learned representations in $\theta$ space to reflect information about site and MR contrasts . 
Figure~\ref{fig:theta_space}(a) shows the learned contrast space of \Tone, \Ttwo, PD-w, and FLAIR images from ten representative sites. 
Each point in the plot corresponds to a 3D MR volume, and the $\theta$ value of each MR volume is calculated by averaging the $\theta$'s of the center 20 axial slices per volume.
The results demonstrate that the four MR contrasts are well-separated in $\theta$ space. 
Furthermore, we observed that the $\theta$ values of PD-w and \Ttwo images are located next to each other, which is consistent with the fact that these two contrasts are often acquired simultaneously with different echo times.

To investigate the impact of sites on the learned $\theta$ values, we plotted the $\theta$ values of \Tone images from the ten sites in Fig.~\ref{fig:theta_space}(b).
The results reveal several interesting observations.
First, the $\theta$ values of images from the same site are generally closer to each other than those from different sites.
Second, images with overlapping $\theta$ clusters share similar acquisition parameters and image contrast, as demonstrated in cases \#4 and \#5 of Figs.~\ref{fig:theta_space}(b) and (c). 
Third, we observed several outliers with $\theta$ values deviating from their main clusters, as showcased by \#1 and \#2 of Fig.~\ref{fig:theta_space}(b).
Upon examination, we discovered that case \#1 is a post-gadolinium \Tone~(post \Tone) image that had erroneous header information that identified it as a pre-gadolinium \Tone~(pre \Tone).
This error is evident from inspection of \#1 in Fig.\ref{fig:theta_space}(c). 
With respect to case \#2, the image was acquired using different parameters than the other images from Site~$S_{12}$.

\begin{figure}[!tb]
    \centering
    \includegraphics[width=0.8\textwidth]{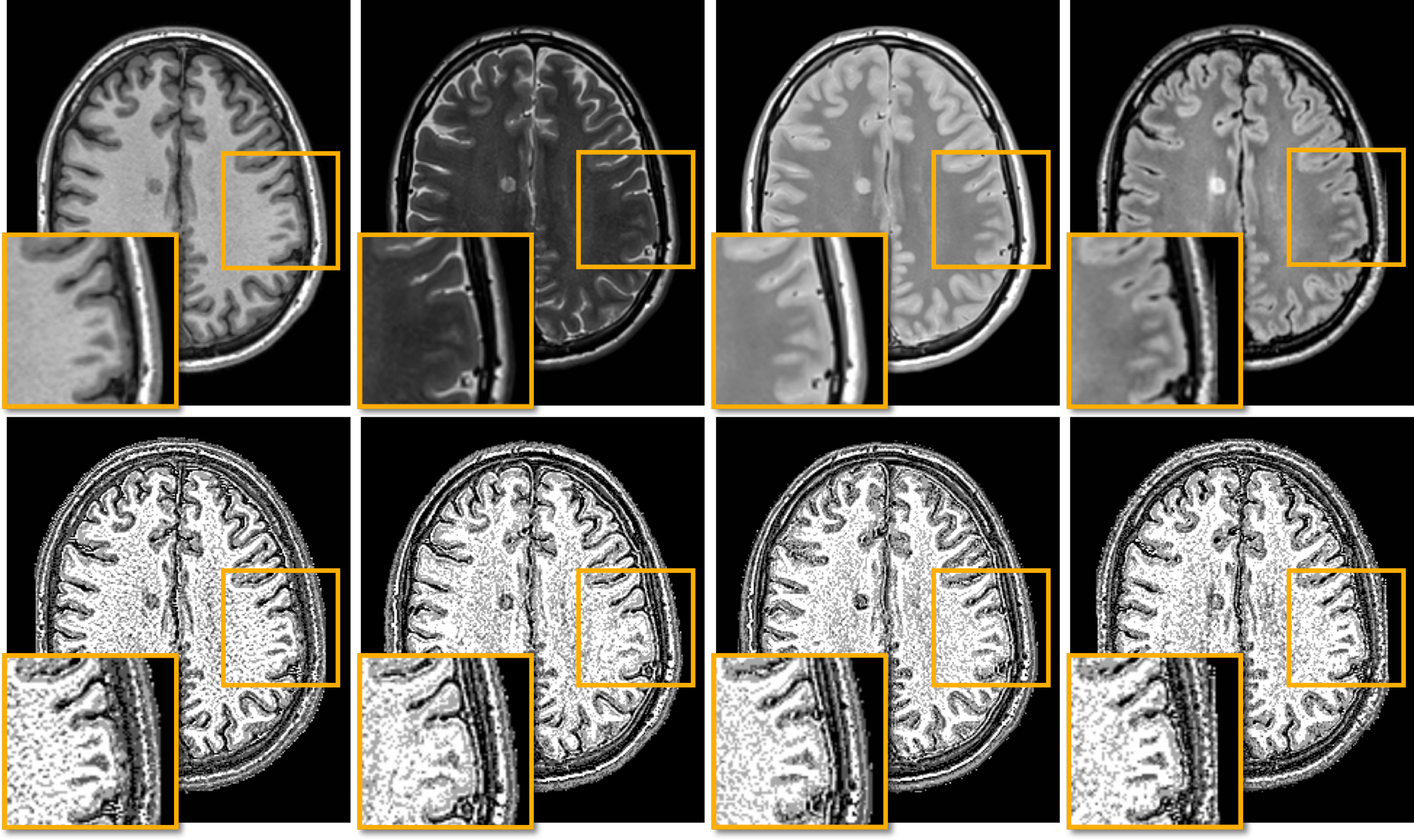}
    \caption{Anatomical representations $\beta$ of intra-site paired data. The top row shows \Tone, \Ttwo, PD-w, and FLAIR images, respectively, with the inset being a zoomed up version of the orange box. The bottom row shows the corresponding $\beta$'s of each contrast and the same zoomed in region.}
    \label{fig:beta_space}
\end{figure}
Figure~\ref{fig:beta_space} shows the learned anatomical representations $\beta$ of intra-site paired \Tone, \Ttwo , PD-w, and FLAIR images.
Generally, $\beta$'s of the four images are visually similar, suggesting that they capture similar anatomical information.
However, there are subtle differences highlighted by the orange boxes, indicating that each MR contrast reveals slightly different anatomical information. 
This observation reassures our motivation behind developing HACA3---different MR contrasts reveal slightly different anatomical information.

\begin{figure}[!tb]
    \centering
    \includegraphics[width=0.99\textwidth]{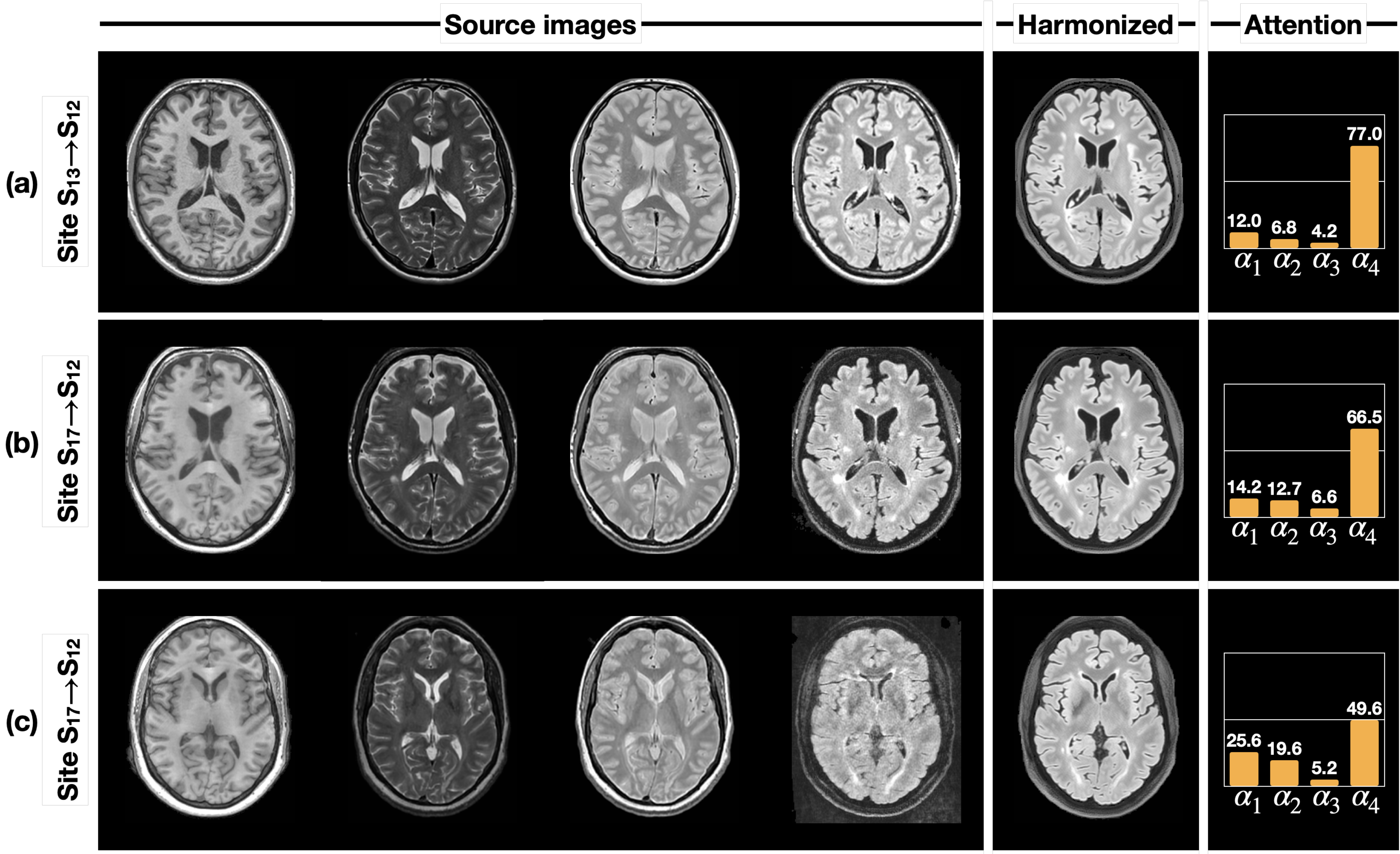}
    \caption{The learned attention $\alpha$ changes with three different harmonization scenarios. In all three scenarios, \Tone, \Ttwo, PD-w, and FLAIR images from sites~($S_{13}$ or $S_{17}$) are harmonized to a FLAIR image of a different site---$S_{12}$ in this case.}
    \label{fig:explore_attention}
\end{figure}
In our previous work~\citep{zuo2023latent}, we have demonstrated that the artifact encoder effectively captures various cases of poor quality images.
However, it is also crucial to ensure that our attention mechanism works properly in highlighting similar contrast source images and downplaying the role of poor quality source images.
To investigate this, we present three harmonization scenarios where \Tone, \Ttwo, PD-w, and FLAIR images from Sites~$S_{13}$ or $S_{17}$ are harmonized to a FLAIR image from Site~$S_{12}$---i.e., Sites~$S_{13}$ or $S_{17}$ are the source and Site~$S_{12}$ is the target. 
Figure~\ref{fig:explore_attention}(a) shows the scenario where all four source modalities have good image quality, resulting in most of the attention ($77\%$) being on the FLAIR image of the source site $S_{13}$---see the attention column in Fig.~\ref{fig:explore_attention}.
This makes sense since $\alpha$ is used to select the corresponding anatomical representation $\beta$ during harmonization.
Figure~\ref{fig:explore_attention}(b) depicts another harmonization scenario where the FLAIR image from the source site ($S_{17}$) has higher noise levels.
Here, the attention on the source FLAIR image has decreased, while the other three MR contrasts have increased. 
The attention model seeks anatomical information from other contrasts to compensate for the lower quality of the source FLAIR. 
As a result, the harmonized FLAIR image has a better quality appearance while preserving anatomical details such as the WM lesions.
Figure~\ref{fig:explore_attention}(c) presents an extreme scenario in which the source FLAIR image has even higher noise levels and motion artifacts. 
In this case, the attention $\alpha$ on the source FLAIR image further decreases to seek alternative anatomical information from other contrasts. 
As a result, the harmonized FLAIR image demonstrates improved image quality and anatomical fidelity. 
It is important to note that the decrease in attention from Figs.~\ref{fig:explore_attention}(b) to~(c) is likely due to differences in image quality rather than contrast, as the images in  Figs.~\ref{fig:explore_attention}(b) and~(c) come from the same source site.

\subsection{Numerical comparisons of multi-site MR image harmonization}
\label{sec:multi_site_harmonization}
\subsubsection{Comparing with supervised and unsupervised harmonization methods}
\label{sec:comparison}

In this experiment, we seek a harmonization model that translates \Tone images from a source site to a target site.
We used a held-out dataset with $12$ subjects traveling across Sites~$S_{11}$~(source) and $S_{12}$~(target) to quantitatively evaluate HACA3 and other methods.
The same traveling dataset was also used in~\citet{dewey2019deepharmony} for evaluation.
These methods come from three broad types: 1)~unsupervised I2I including CycleGAN~\citep{zhu2017unpaired} and CUT~\citep{park2020contrastive},
2)~two unsupervised harmonization methods based on intra-site paired data~\citep{adeli2021representation,zuo2021unsupervised}, and 3)~supervised harmonization~\citep{dewey2019deepharmony}. 
Structural similarity index~(SSIM)~\citep{wang2004image} and peak signal-to-noise ratio~(PSNR) are used to quantitatively evaluate all methods. 
Throughout the paper, both SSIM and PSNR are calculated on the entire MR volume.

\begin{figure}[!tb]
    \centering
    \includegraphics[width=0.99\textwidth]{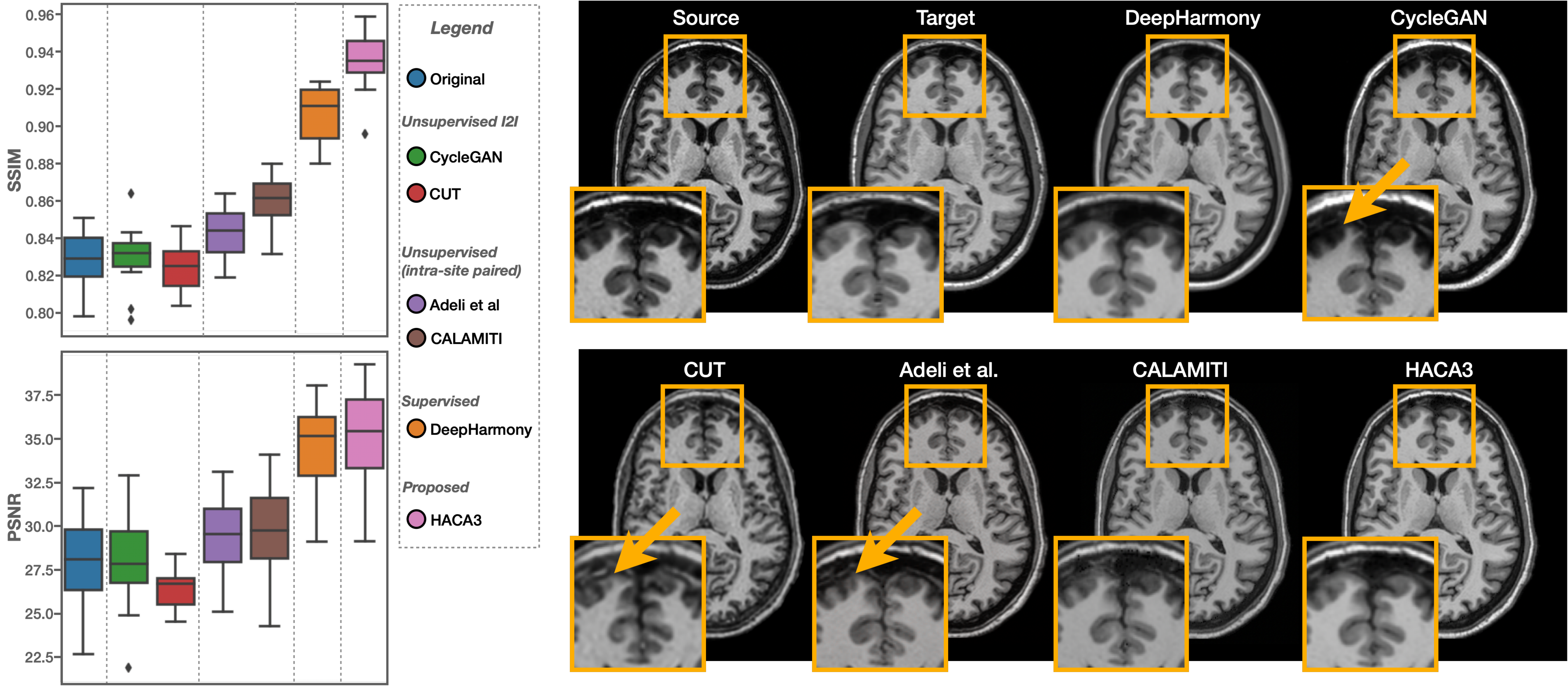}
    \caption{Numerical comparisons between HACA3 (proposed) and other methods using a held-out dataset of inter-site traveling subjects. SSIM and PSNR of \Tone images are calculated. Example \Tone images are shown on the right.}
    \label{fig:ssim_psnr}
\end{figure}

\textit{Compared with unsupervised I2I with cycle consistency constraint in anatomy:} 
As shown in Fig.~\ref{fig:ssim_psnr}, we compared HACA3~(pink) with CycleGAN~(green) and CUT~(red).
Both CycleGAN and CUT were trained on unpaired \Tone images from Sites~$S_{11}$ and $S_{12}$. 
HACA3 outperforms both methods with statistical significance ($p<0.01$ in a paired Wilcoxon signed-rank test). 
Surprisingly, CycleGAN and CUT did not show much improvement compared to the unharmonized images~(blue), even though the synthetic images are visually fine, as shown in Fig.~\ref{fig:ssim_psnr}.
We hypothesize that this may be due to the issue of geometry shift, as indicated by the orange arrows in Fig.~\ref{fig:ssim_psnr}.
This observation reassures the findings in previous studies~\citep{gebre2023cross, yang2018unpaired} that the cycle consistency constraint for anatomy alone is not sufficient for MR harmonization.

\textit{Compared with unsupervised harmonization based-on intra-site paired data:} 
We then compared HACA3 to existing unsupervised harmonization methods that are also based on intra-site paired data, such as Adeli~et~al.~\citep{adeli2021representation} and CALAMITI~\citep{zuo2021unsupervised}.
Both methods were trained on intra-site paired \Tone and \Ttwo images with disentanglement.
As shown in Fig.~\ref{fig:ssim_psnr}(a), HACA3~(pink) outperforms these methods~(purple and brown) with statistical significance ($p<0.01$ in a paired Wilcoxon signed-rank test). 
Given that all three methods are based on intra-site paired images for training, we believe that the superior performance of HACA3 comes from its ability to use multiple MR contrasts during application.
In Sec.~\ref{sec:ablation}, we further explore the impact of this ability with various cases of input MR contrasts.
Interestingly, all three methods have better performance than unsupervised I2I methods, which demonstrates the benefits of using intra-site paired data in harmonization.

\textit{Compared with supervised harmonization:}
Given that HACA3 can be trained on a wide variety of data, we finally ask ourselves whether it could potentially outperform supervised harmonization methods.
To test this hypothesis, we compared HACA3 with DeepHarmony~\citep{dewey2019deepharmony}, a supervised harmonization method that was specifically trained on inter-site paired images from $S_{11}$ and $S_{12}$.
The same evaluation dataset as \cite{dewey2019deepharmony} was used here to evluate HACA3 and other comparison methods. 
Paired Wilcoxon signed rank test shows that HACA3 outperforms DeepHarmony~(orange) in SSIM~(see Fig.~\ref{fig:ssim_psnr}) with statistical significance ($p< 0.01$).
This result highlights the potential of HACA3 as a versatile and effective harmonization method.

\subsubsection{Ablation: HACA3 handling missing contrasts}
\label{sec:ablation}
\begin{figure}[!tb]
    \centering
    \begin{tabular}{cc}
        \includegraphics[width=0.45\textwidth]{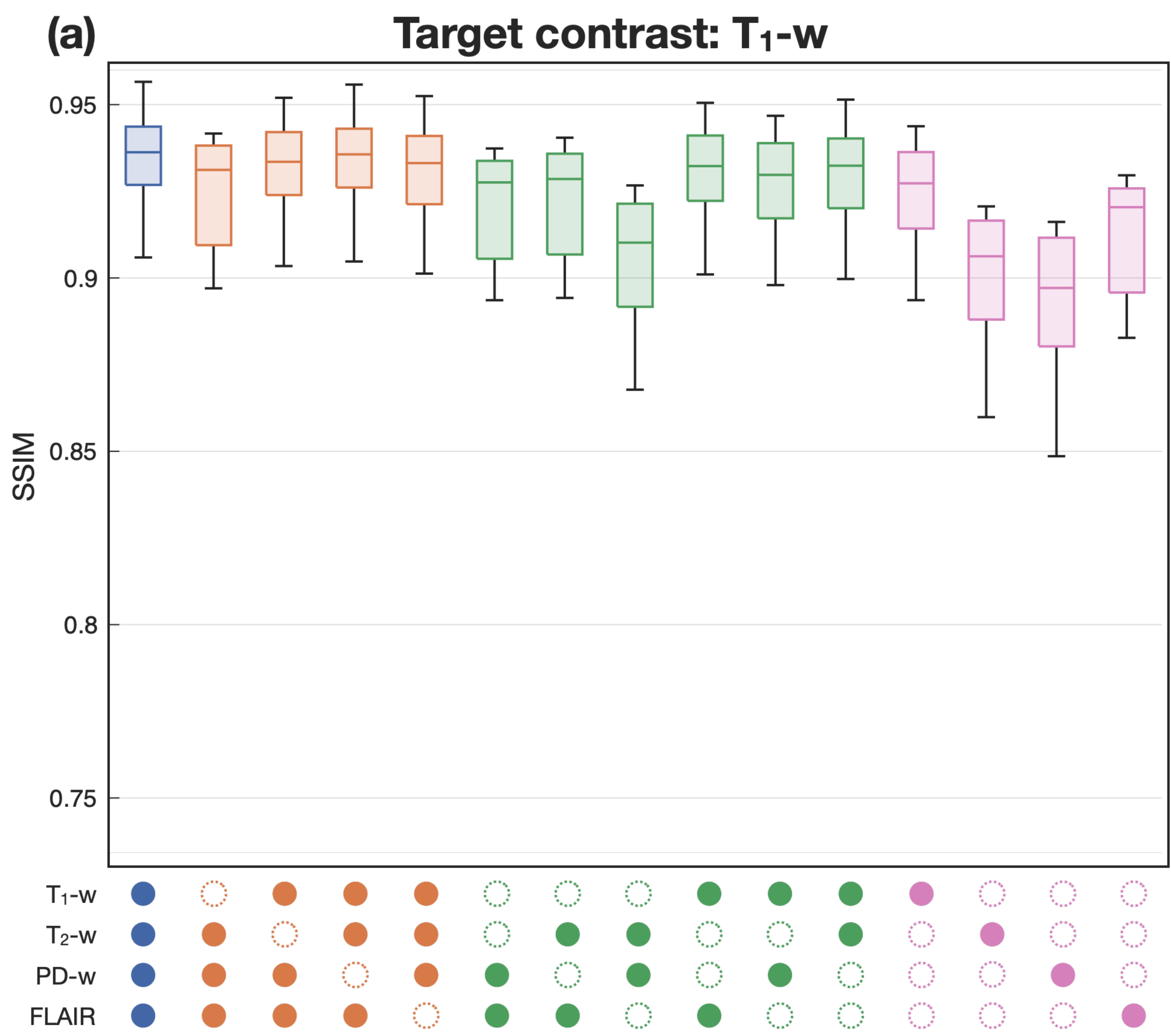} & 
        \includegraphics[width=0.45\textwidth]{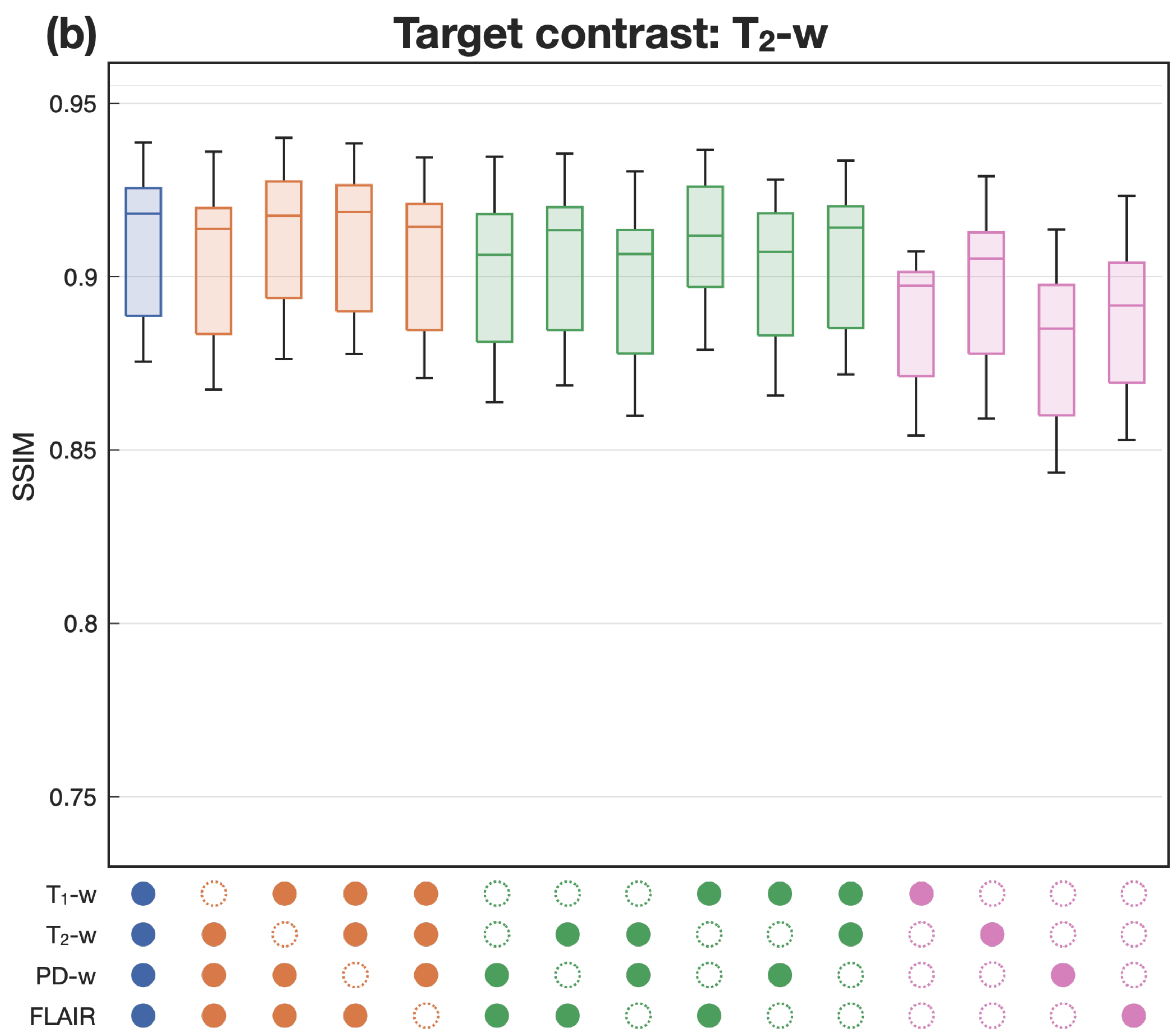} \\
        & \\
        \includegraphics[width=0.45\textwidth]{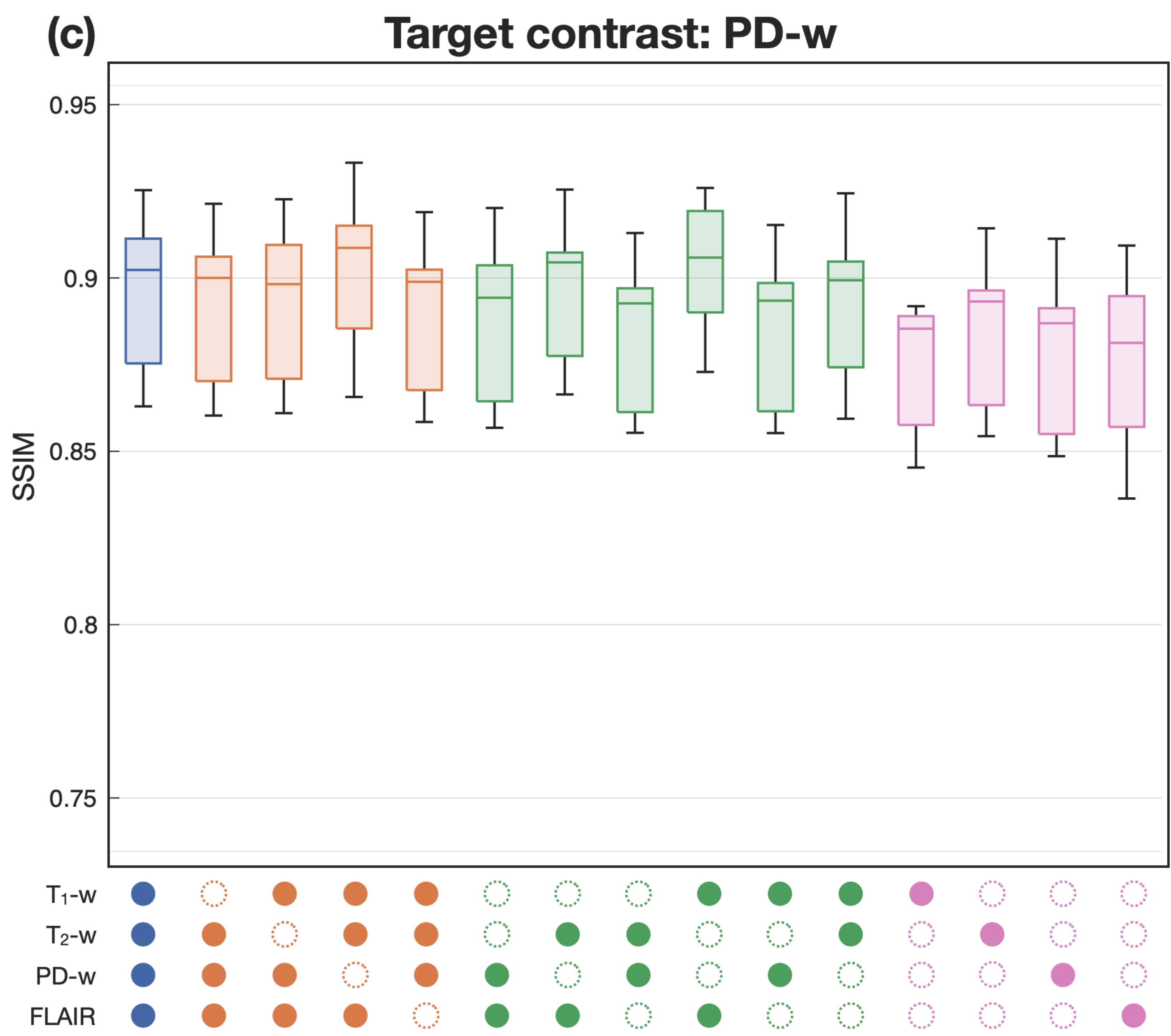} & 
        \includegraphics[width=0.45\textwidth]{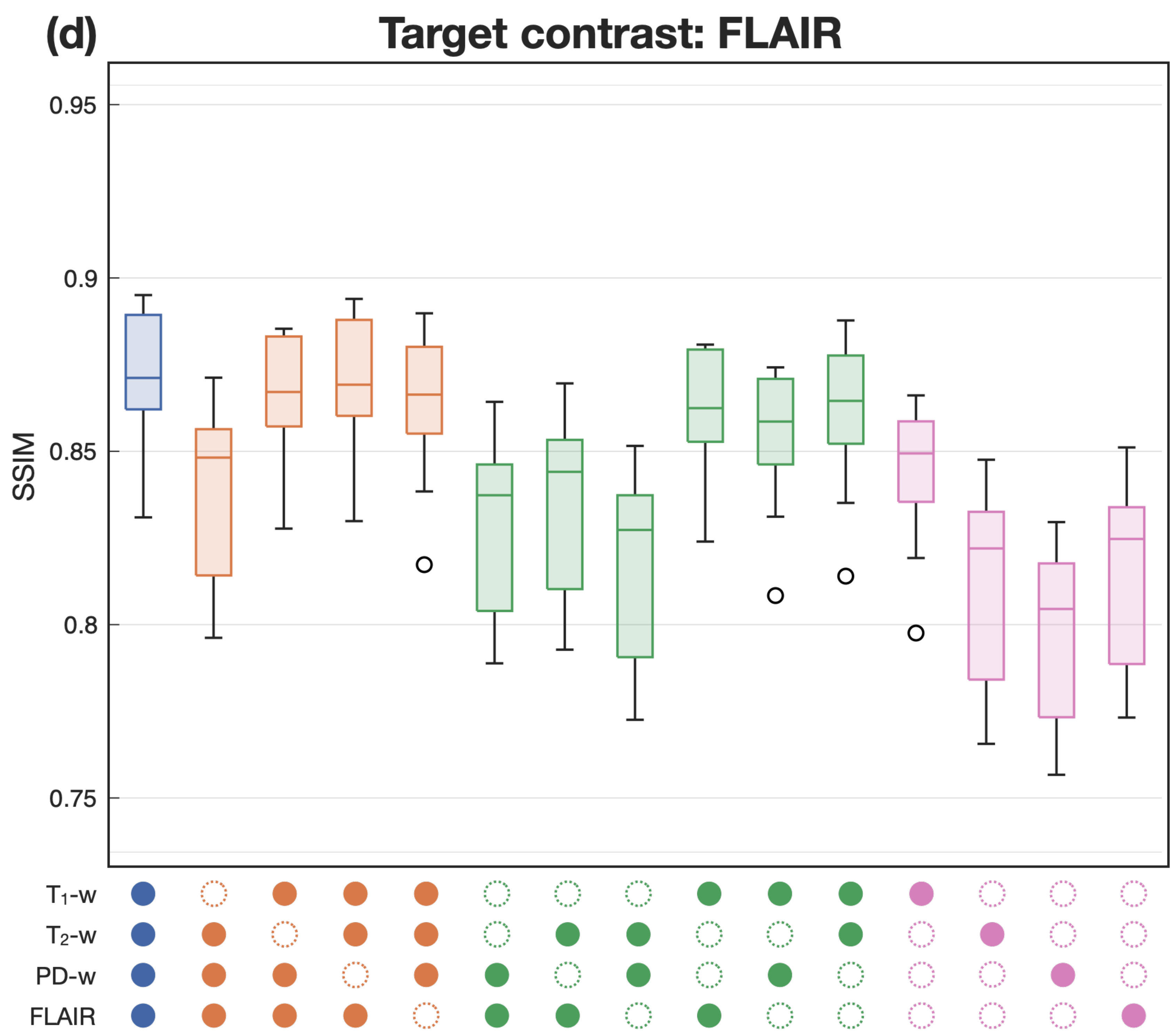} 
    \end{tabular}
    \caption{HACA3 handling different availability of source images. From \textbf{(a)} to \textbf{(d)}: target image being \Tone, \Ttwo, PD-w, and FLAIR images, respectively. Colored boxplots represent different numbers of source images. The panel below the boxplots indicates which images were used as input to the harmonization (with an empty circle indicating the absence of a particular contrast). }
    \label{fig:ablation}
\end{figure}

HACA3 is designed to handle any number of source MR contrasts during both training and application.
To investigate this ability and the impact of each source contrast on the final harmonization result, we conducted an ablation study on all possible scenarios during application.
Specifically, we used the same inter-site traveling dataset as Sec.~\ref{sec:comparison} in our ablation study~($N=12$ subjects traveled between Sites~$S_{11}$ and $S_{12}$ with \Tone, \Ttwo, PD-w, and FLAIR images).
We then applied HACA3 to harmonize images from $S_{11}$ to $S_{12}$ and reported the SSIM values between the harmonized image and the real $S_{12}$ image for each of \Tone, \Ttwo, PD-w, and FLAIR being the target contrast.
Results in Figs.~\ref{fig:ablation}(a)--(d) demonstrate HACA3's robust performance across various combinations of input contrasts. 
When all four contrasts are used as input, as well as when only three or two are available, the results in terms of SSIM are generally similar. 
However, having all four contrasts available typically results in the best performance.
For our target contrasts of \Tone, \Ttwo, and FLAIR, we observed that missing the corresponding source image often has an negative impact on the harmonization results.
However, when PD-w is the target contrast, the performance deviates from this pattern. 
In this case, missing PD-w as the source image actually improves the results.
We hypothesize that this is due, in part, to the lower resolution of PD-w and \Ttwo images.
Lastly, when FLAIR is the target contrast, the harmonization performance is relatively lower compared to the other target contrasts.
This can be attributed to the challenges in synthesizing WM lesions, which are harder to replicate accurately.
HACA3 heavily relies on information from \Tone and FLAIR images to achieve this task. 
Overall, our study highlights the robustness of HACA3 in handling various input contrasts and sheds light on the factors influencing its performance.

\subsection{Evaluating HACA3 in downstream tasks}
\label{sec:downstream_tasks}
To validate HACA3's ability to alleviate domain shift, we showcase two different downstream image analysis tasks: 1)~WM lesion segmentation and 2)~whole brain parcellation. 
The first task is based on multi-site cross-sectional data and the second task focuses on longitudinal analyses with scanner change and upgrades.

\subsubsection{WM lesion segmentation on multi-site data} 
\begin{figure}[!tb]
    \centering
    \begin{tabular}{ccc}
        \includegraphics[width=0.29\columnwidth]{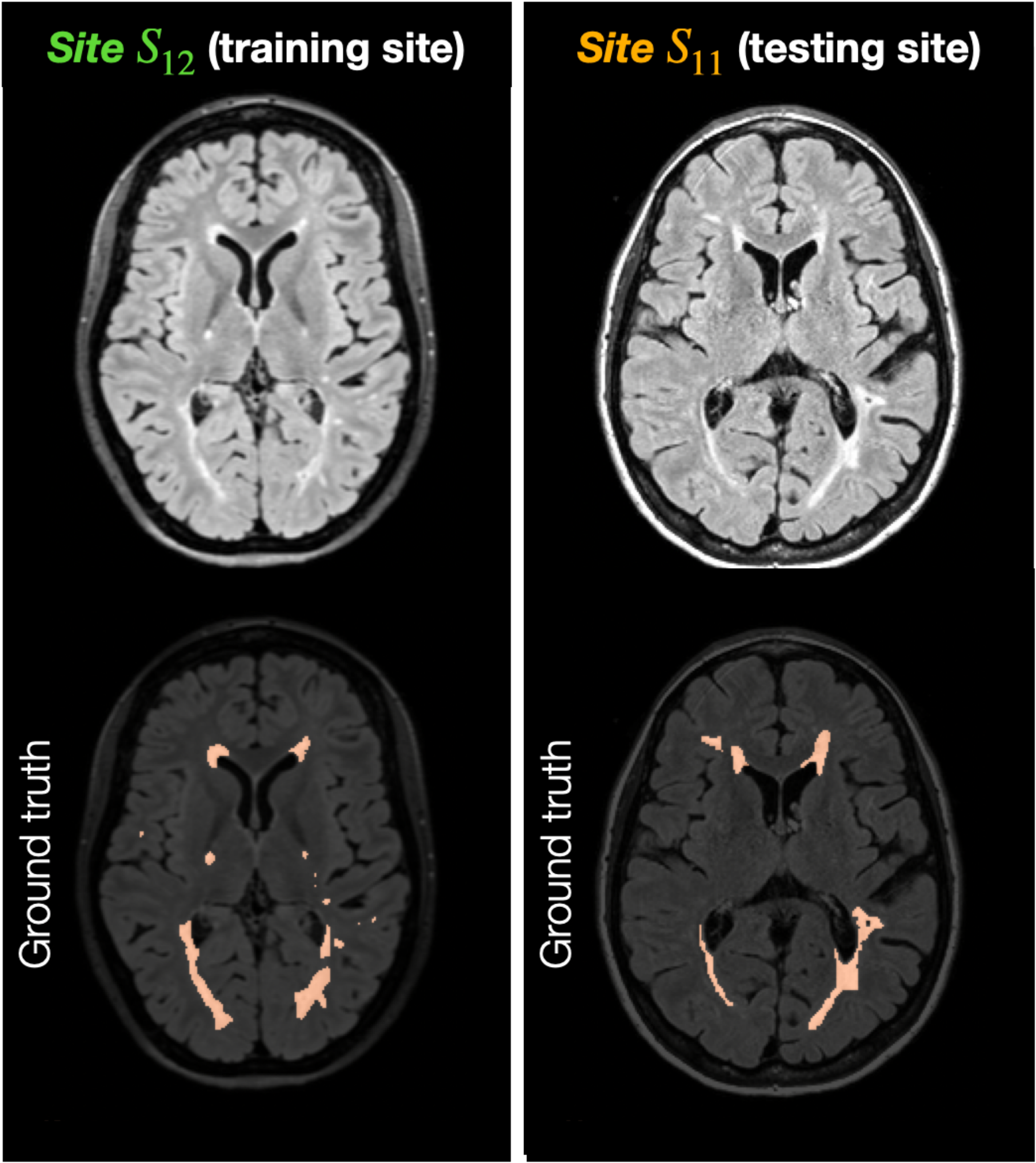} &~~&
         \includegraphics[width=0.62\columnwidth]{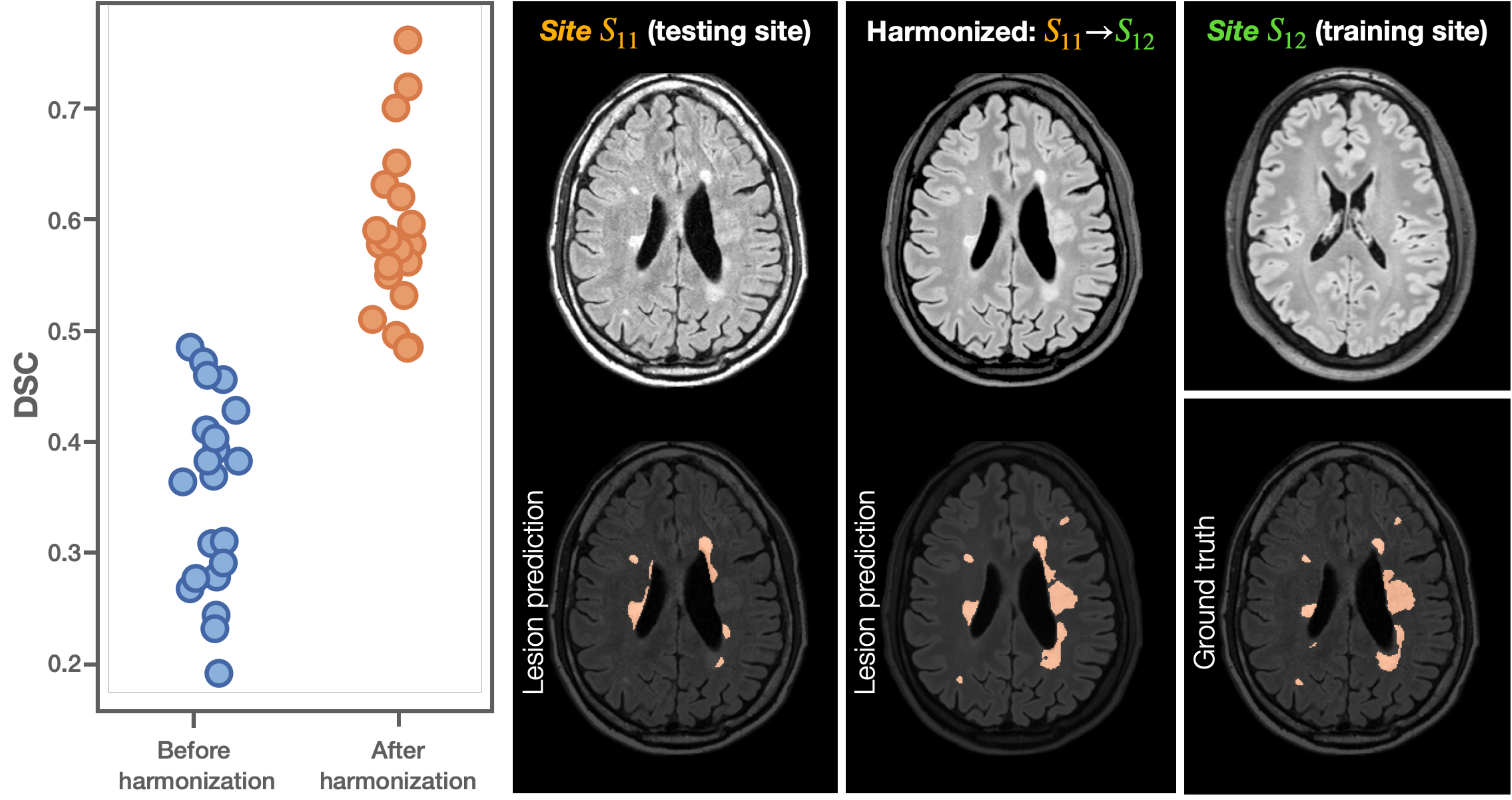} \\
         \textbf{(a)} & & \textbf{(b)}
    \end{tabular}
    \caption{\textbf{(a)} Training and testing sites for WM lesion segmentation with a 3D U-Net. \textbf{(b)}~DSC showed improvements after harmonizing images from the testing site~(Site $S_{11}$) to the lesion training site~(Site $S_{12}$). Example images are shown on the right.} 
    \label{fig:lesion_segmentation}
\end{figure}

As shown in Fig.~\ref{fig:lesion_segmentation}(a), two MS datasets acquired from sites~$S_{11}$ and $S_{12}$ were used in this experiment. 
The training data~($S_{12}$) for MS lesion segmentation include \Tone~(not shown), FLAIR, and expert delineations of WM lesions of $10$ subjects. 
The testing data~($S_{11}$) to evaluate lesion segmentation come from ILLSC 2015~\citep{carass2017longitudinal}, which is publicly available.
A 3D U-Net with four downsamplings was trained with MR images and delineations from $S_{12}$ using a Dice similarity coefficient~(DSC) loss.

The 3D U-Net achieved a DSC of $0.593 \pm 0.072$ (similar to the best results reported in~\cite{tohidi2022multiple} and close to the inter-rater variability of~\cite{carass2017longitudinal}) in a five-fold cross validation on $S_{12}$, which it was trained on.
However, when the 3D U-Net was applied to $S_{11}$, the DSC dropped to $0.348 \pm 0.089$ due to domain shift.
HACA3 was then applied to harmonize images from Site~$S_{11}$ to $S_{12}$ aiming at alleviating domain shift, and the lesion segmentation was reevaluated. 
As shown in Fig.~\ref{fig:lesion_segmentation}(b), DSC has improved to $0.590 \pm 0.075$, which is similar to the performance on the training site.
It is worth noting that WM lesions are particularly difficult to synthesize and characterize due to the large variation in lesion size and location. 
It is encouraging that HACA3 generates high fidelity images that show effectiveness in WM lesion segmentation both qualitatively and quantitatively. 

\subsubsection{Whole brain parcellation on longitudinal data.}
\begin{table}[!tb]
    \centering
    \caption{Longitudinal ICCs and $\sigma_\epsilon^2$ of cGM, WM, and LatV before and after harmonization. Details about each dataset are shown in Table~\ref{tab:dataset} (Sites $S_3$ to $S_{10}$). For longitudinal ICC higher values are better, while for $\sigma_\epsilon^2$ lower values are better.}
    \label{tab:icc}
    \adjustbox{max width = 0.99\textwidth}{
    \begin{tabular}{c c c c c c c c c}
        \toprule
        \multirow{2}{*}{\textbf{Dataset}} & \multirow{2}{*}{\textbf{\# Subjects}} &
        \multirow{2}{*}{\textbf{\# Sessions}} & \multirow{2}{*}{\textbf{Structure}} & \multicolumn{2}{c}{\textbf{ICC (\%)}} & \multicolumn{2}{c}{$\sigma_\epsilon^2$} \\
        \cmidrule(lr){5-6} \cmidrule(lr){7-8} 
        {} & {} & {} & {} & {Before} & {After} & {Before} & {After} \\
        \midrule 
        \multirow{3}{*}{\texttt{OASIS3}} & \multirow{3}{*}{721} & \multirow{3}{*}{1,117} & {cGM} & $81.95$ & $\boldsymbol{95.13}$ & $83.6$ & $\boldsymbol{44.8}$  \\
        {} & {} & {} & {WM} & $83.54$ & $\boldsymbol{95.85}$ & $64.1$ & $\boldsymbol{31.9}$  \\
        {} & {} & {} & {LatV} & $96.37$ & $\boldsymbol{96.38}$ & $25.4$ & $\boldsymbol{25.2}$  \\
        \midrule
        \multirow{3}{*}{\texttt{BLSA}} & \multirow{3}{*}{1,037} & \multirow{3}{*}{2,655} & {cGM} & $86.98$ & $\boldsymbol{96.49}$ & $106.9$ & $\boldsymbol{52.1}$  \\
        {} & {} & {} & {WM} &  $87.35$ & $\boldsymbol{96.38}$ & $133.1$ & $\boldsymbol{59.3}$  \\
        {} & {} & {} & {LatV} & $95.96$ & $\boldsymbol{95.99}$ & $46.2$ & $\boldsymbol{29.7}$  \\
        \bottomrule
    \end{tabular}}
\end{table}
We used two public longitudinal datasets, i.e., OASIS3~(Sites~$S_3$ to $S_6$)~\citep{LaMontagne} and BLSA~(Sites~$S_7$ to $S_{10}$)~\citep{resnick2000one}, to evaluate HACA3 for longitudinal analyses. 
The number of subjects and sessions of each dataset is shown in Table~\ref{tab:icc}.
The same preprocessing was applied here, followed by a whole brain parcellation on \Tone images using~\cite{huo20193d}.
For the cortical GM~(cGM), cerebral WM~(WM), and lateral ventricles~(LatV), a structure-specific
linear mixed effects~(LME) model $y_{ij} = a_0 + a_1 x_{ij} + b_j + \epsilon_{ij}$ was fitted, 
where $x_{ij}$ and $y_{ij}$ are age and percentage structural volume~(structural volume divided by total brain volume)
of session $i$ and subject $j$, respectively. 
We reuse the notations $x$, $y$, $i$, and $j$ to be consistent with the LME literature~\citep{erus2018longitudinally}. 
$b_j \sim \mathcal{N}(0,\sigma_b^2)$ is the subject-specific bias, and $\sigma_b^2$ models population variance.
$\epsilon_{ij} \sim \mathcal{N}(0, \sigma_\epsilon^2)$ is the error term modeling noise in observations.
Based on the LME, longitudinal intra-class correlations~(ICCs) were calculated to characterize the effect of harmonization in longitudinal analysis with ICC defined by, 
\begin{equation*}
    \text{ICC} = \frac{\sigma_b^2}{\sigma_b^2 + \sigma_\epsilon^2},
\end{equation*}
where an ICC close to $0\%$ means the noise in observations is the dominant factor over population difference.
An ICC close to $100\%$ indicates most variances are due to the natural population difference rather than noisy observations. 
Assuming the effect of scanner change and upgrades are alleviated with harmonization, we would expect increased ICCs after harmonization.
Table~\ref{tab:icc} shows that the ICCs and $\sigma_\epsilon^2$ of all structures from both datasets were improved after harmonization. 

\section{Discussions and Conclusion}
\label{sec:discussion_and_conclusion}
In this paper, we present HACA3, a novel harmonization approach with attention-based contrast, anatomy, and artifact awareness. 
We demonstrate the effectiveness of HACA3 through extensive experiments and evaluations on diverse MR datasets.
HACA3 successfully learns a disentangled latent space of contrast and anatomy, allowing different MR contrasts and imaging sites to be differentiated in the contrast space $\theta$.
This demonstrates HACA3's capability to capture complex information about image acquisition and contrast, which is crucial for contrast-accurate MR image harmonization.
Moreover, we show that the anatomical representations $\beta$ of intra-site paired images, while generally similar, reveal slight different anatomical features. 
This finding is consistent with HACA3's design, which respects the inherent anatomical differences between MR contrasts.
HACA3's capability to understand these nuanced anatomical features is essential for generating harmonized images with high anatomical fidelity. 
The learned artifact representations $\eta$, not only inform HACA3 for robust harmonization, but also provide rich information for MR quality control, as we have demonstrated in another work~\citep{zuo2023latent}. 
Our attention mechanism based on $\eta$ and $\theta$ effectively identifies poor quality images at the source site and learns to dynamically combine anatomical information.

By respecting contrast and artifacts, HACA3 ensures that the harmonized images are high quality and suitable for downstream image analyses. Numerical comparisons show that HACA3 significantly outperforms unsupervised I2I methods, unsupervised harmonization methods based on intra-site paired images, and supervised harmonization methods. 
We have also explored the impact of different source image availability on harmonization results, demonstrating HACA3's robustness under varying input conditions. 
Our study highlights the potential of HACA3 to alleviate domain shift in neuroimage analyses through two different downstream tasks. 
In the WM lesion segmentation task, HACA3 provides high-quality synthetic FLAIR images with preserved lesion structure.
For the longitudinal volumetric analysis task, HACA3 promotes consistent longitudinal volumetric analyses in terms of longitudinal ICCs and error residual.

Despite its strengths, we discuss some limitations and intriguing findings that may motivate future research. 
First, the imbalance of available MR contrasts in training data may have a negative impact on harmonization performance. 
Specifically, if a contrast is missing in many training sites, HACA3 may not be sufficiently trained due to the lack of training data of that particular contrast. 
It is worth noting that this issue of imbalanced training data is not specific to HACA3, but is present in most deep learning methods. 
We believe this issue can be mitigated by importance sampling training data based on the prevalence of each contrast, so each MR contrast would appear equally during training.
Second, HACA3 currently focuses on \Tone, \Ttwo, PD-w, and FLAIR images.
While this covers a large range of MR contrasts in clinical applications, we believe HACA3's capability is beyond that. 
As we show in Fig.~\ref{fig:theta_space}, even though HACA3 was not trained on post \Tone, the theta-encoder is still able to capture this difference in contrast.
This capacity should be explored in future research for potential extension of HACA3 to include more MR contrasts and even other imaging modalities (e.g., computational tomography~\citep{chartsias2019disentangled}).
Third, the attention $\alpha$ can be extended to incorporate spatial variations. 
Current attention mechanism in HACA3 is global, meaning that each MR contrast receives a single number during anatomy fusion. 
By allowing the attention to vary across spatial locations, it can better adapt to the local anatomical features and provide more fine-grained control over the harmonization process.
This could potentially lead to better performance, especially in areas with complex or subtle anatomical differences, depending on the source images and target contrasts.

In conclusion, our work on HACA3 showcases its ability to address challenges in MR image harmonization and its potential to improve the quality and consistency of neuroimaging studies. 
By successfully disentangling contrast and anatomy, respecting inherent anatomical differences, and leveraging attention mechanisms for handling artifacts, HACA3 sets a new benchmark in MR image harmonization and promises to advance the field of harmonization. 
Future research should focus on addressing the limitations and further expanding the applicability of HACA3 to a wider range of MR contrasts and imaging scenarios.

\section*{Acknowledgements}
The authors thank BLSA participants, as well as colleagues of the Laboratory of Behavioral Neuroscience and the Image Analysis and Communications Laboratory.
This work was supported in part by the Intramural Research Program of the National Institutes of Health, National Institute on Aging and in part by the TREAT-MS study funded by the Patient-Centered Outcomes Research Institute~(PCORI) grant MS-1610-37115~(Co-PIs: Drs. S.D. Newsome and E.M. Mowry).

\printcredits

\bibliographystyle{cas-model2-names}

\bibliography{cas-refs}





\end{document}